\documentclass[aps,preprint,floatfix]{revtex4}%
\usepackage{amsfonts}
\usepackage{amsmath}

\usepackage{epsfig}
\usepackage{amssymb}

\usepackage[figuresright]{rotating}
\usepackage{color}

\newcommand{\kval}{\kappa_{\rm val}}
\newcommand{\ksea}{\kappa_{\rm sea}}

\newcommand{\be}{\begin{equation}}
\newcommand{\ee}{\end{equation}}
\newcommand{\ba}{\begin{array}}
\newcommand{\ea}{\end{array}}
\newcommand{\bea}{\begin{eqnarray}}
\newcommand{\eea}{\end{eqnarray}}

\newcommand{\err}[2]{${\scriptstyle {}^{+{#1}}_{-{#2}}}$}
\newcommand{\er}[2]{{\scriptstyle {}^{+{#1}}_{-{#2}}}}

\begin{document}
\begin{flushright}
Argonne Preprint PHY-12182-TH-2008 \\
CSSM Preprint ADP-08-10/T670 \\
JLab Preprint JLAB-THY-08-884 \\
\end{flushright}
\vspace*{3mm}
\title[Chiral Extrapolation of pQQCD]{An analysis of the nucleon
spectrum from lattice partially-quenched QCD}

\author{W.~Armour}
\author{C.~R.~Allton}
\affiliation{Department of Physics, Swansea University, Swansea,
SA2~8PP, Wales, U.K.}
\author{D.~B.~Leinweber}
\affiliation{
Special Research Centre for the Subatomic Structure of Matter (CSSM),
School of Chemistry \& Physics, University of Adelaide 5005, Australia
}
\author{A.~W.~Thomas}
\affiliation{Jefferson Lab, 12000 Jefferson Ave., Newport News, VA
  23606, USA}
\affiliation{College of William and Mary, Williamsburg, VA 23187, USA}
\author{R.~D.~Young}
\affiliation{Physics Division, Argonne National Laboratory, Argonne, IL 60439,
USA}

\begin{abstract}

The chiral extrapolation of the nucleon mass, $M_n$, is investigated
using data coming from 2-flavour partially-quenched lattice
simulations.  A large sample of lattice results from the CP-PACS
Collaboration is analysed using the leading one-loop corrections, with
explicit corrections for finite lattice spacing artifacts. The
extrapolation is studied using finite-range regularised chiral
perturbation theory.  The analysis also provides a quantitative
estimate of the leading finite volume corrections. It is found that
the discretisation, finite volume and partial quenching effects can
all be very well described in this framework, producing an
extrapolated value of $M_n$ in agreement with experiment.
Furthermore, determinations of the low energy constants of the nucleon
mass's chiral expansion are in agreement with previous methods, but
with significantly reduced errors.  This procedure is also compared
with extrapolations based on polynomial forms, where the results are
less encouraging.

\end{abstract}
\startpage{1}
\endpage{2}
\maketitle



\section{Introduction}

There has been great progress in lattice QCD in recent years,
associated both with Moore's Law and with improved algorithms, which
mean that one can work with larger lattice spacings and still
approximate the continuum limit well. The CP-PACS group has devoted
considerable effort to the study of the masses of the lowest mass
baryons and vector mesons.  This has led, for example, to a comprehensive set
of data for the mass of the nucleon in partially-quenched QCD (pQQCD),
with exceptionally small statistical errors~\cite{cppacs}.  We shall
exploit this data.

The remaining barrier to direct comparison with experimental data is
the fact that calculations take much longer as the quark mass
approaches the chiral limit. Indeed the time for a given calculation
scales somewhere in the range $m_\pi^{-4}$ to $m_\pi^{-9}$, depending
on how hard one works to preserve chiral symmetry
\cite{Jansen:2008vs}. As a result there
has been considerable interest in using chiral perturbation theory
($\chi$PT), an effective field theory (EFT) built on the symmetries of
QCD, to provide a functional form for hadron properties as a function
of quark mass~\cite{Leinweber:2003dg,Procura:2003ig,Leinweber:2001ui}.  
In principle,
such a functional form can then be used to extrapolate from the large
pion masses where lattice data exists to the physical
value. Unfortunately, there is considerable evidence that the
convergence of dimensionally regularised $\chi$PT is too slow for this
expansion to be reliable at present
\cite{Leinweber:1999ig,Bernard:2002yk,Young:2002ib,Durr:2002zx,%
Beane:2004ks,Thomas:2004iw,Leinweber:2005xz,McGovern:2006fm}.

On the other hand, it can be shown that a reformulation of $\chi$PT
using finite-range regularisation (FRR) effectively re-sums the chiral
expansion, leaving a residual series with much better convergence
properties \cite{Leinweber:2003dg,Young:2002ib}.  The FRR expansion is
mathematically equivalent to dimensionally regularised $\chi$PT to the
finite order one is working \cite{Donoghue:1998bs,Young:2002ib}.
Systematic errors associated with the functional form of the regulator
are at the fraction of a percent level \cite{Leinweber:2003dg}.
A formal description of the formulation of baryon $\chi$PT using a
momentum cutoff (or FRR) have recently been considered by Djukanovic
{\it et al.}~\cite{Djukanovic:2005jy}.
The price of such an approach is a residual dependence on the
regulator mass, which governs the manner in which the loop integrals
vanish as the pion mass grows large.  However, if it can be
demonstrated that reasonable variation of this mass does not
significantly change the extrapolated values of physical properties,
one has made progress.  This seems to be the case for the nucleon mass
\cite{Thomas:2004iw} and magnetic moments \cite{Young:2004tb}, for
example, where ``reasonable variation'' is taken to be $\pm$20\%
around the best fit value of the regulator mass.

In order to test whether the problem is indeed solved in this way one
needs a large body of accurate data. This is in fact available for the
nucleon, where CP-PACS has carried out lattice simulations of pQQCD
with a wide range of sea and valence
masses. This sector requires a modified effective field theory, namely
partially-quenched chiral perturbation theory (pQ$\chi$PT)
\cite{Golterman:1997st,Sharpe:2001fh}. Formal developments in this
field have made significant progress in the study of a range of
hadronic observables --- see
Refs.~\cite{Chen:2001yi,Beane:2002vq,Leinweber:2002qb,Arndt:2003ww,%
Arndt:2004bg,Bijnens:2004hk,Detmold:2005pt}, for example.

This large body of pQQCD simulation data is analysed within a
framework which incorporates the leading low energy behaviour of
partially-quenched EFT. Finite-range regularisation is implemented
to evaluate loop integrals, for reasons discussed above.  The aim is
to test whether this approach produces a more satisfactory description
of the complete data set than the more commonly used, naive
extrapolation formulas.

As we will see, the finite-range regularisation method is able to
reproduce the nucleon mass with a remarkable level of accuracy using
partially-quenched lattice data at only relatively large pion
mass. 
Furthermore we are able to determine the low energy constants of the 
chiral expansion of the nucleon mass to a remarkable level of accuracy.
For this reason, we encourage the generation of
partially-quenched data by the lattice community since it greatly
increases the coverage of parameter space, thus enabling chiral
extrapolations to the physical point to be performed more accurately.

This work is a companion paper to \cite{rhopapers} in which we used
the same technique to analyse the vector meson mass, obtaining the
$\rho-$meson mass to 1\% of its physical value.

The next section summarises the finite-range regularised forms for the
self-energy of the nucleon in the case of pQQCD.
Section~\ref{sec:cppacs} discusses the data used from 
the CP-PACS Collaboration~\cite{cppacs}.
We then give details of the chiral fits in Sec.~\ref{sec:fits}.
Finally,
Sec.~\ref{sec:expt} reports the consequences of the fits for the   
determination of the
nucleon mass at the physical point.  



\section{Self-Energies for the Partially-Quenched Analysis}
\label{sec:se}
Theoretical calculations of dynamical-fermion QCD provide an
opportunity to explore the properties of QCD in an expansive manner.
The idea is that the sea quark masses (considered in generating the
gauge fields of the QCD vacuum) and valence quark masses (associated
with operators acting on the QCD vacuum) need not match.  Such
simulation results are commonly referred to as partially-quenched
calculations.  Unlike quenched QCD, which connects to full QCD only in
the heavy quark limit, pQQCD is not an approximation.
The chiral coefficients of terms in the chiral expansion (such as the
axial couplings of the $\pi$ and $\eta'$) are the same as in full QCD.
Hence, the results of pQQCD provide a theoretical
extension of QCD \cite{Sharpe:2001fh}. QCD, as realized in nature, is
recovered in the limit where the valence and sea masses match.

In this section we explain the form of the finite-range regularised
chiral extrapolation formula in the case of pQQCD
--- i.e., the case where the valence and sea quarks are not
necessarily mass degenerate.  This work extends on the early work of
Ref.~\cite{Leinweber:1993yw}, and mirrors our analysis of the vector
meson mass in Refs.~\cite{rhopapers}.

We restrict our attention to correlators of nucleons containing three
degenerate valence quarks. However, loop diagrams can (and do) contain
baryons which contain non-degenerate quarks.  For convenience, we
introduce the following notation for baryon and pseudoscalar meson masses,
$M_{B}(\beta,\ksea;\kval^1,\kval^2,\kval^3)$ and
$M_{PS}(\beta,\ksea;\kval^1,\kval^2)$, where the first two arguments
refer to the gauge coupling and sea quark mass, and the arguments
after the semi-colon refer to the valence quark values.  Throughout
the paper it will be convenient to abbreviate this by
introducing the notation: \bea
\begin{array}{rcl}
M_{B}^{deg} &=& M_{B}(\beta,\ksea;\kval,\kval,\kval) \\
M_{B}^{non\_deg} &=& M_{B}(\beta,\ksea;\ksea,\kval,\kval) \\
M_{PS}^{deg} &=& M_{PS}(\beta,\ksea;\kval,\kval) \\
M_{PS}^{non\_deg} &=& M_{PS}(\beta,\ksea;\ksea,\kval) \\
M_{PS}^{unit} &=& M_{PS}(\beta,\ksea;\ksea,\ksea) 
\end{array}
\eea where the superscript {\em unit} refers to the unitary data;
{\em deg} refers to a hadron containing degenerate valence quarks;
and {\em non-deg} refers to the case where the valence quarks are 
not degenerate.

The derivation of the pQQCD chiral expansion can be described
by diagrammatic methods~\cite{Leinweber:2002qb}, where the role of
sea quark loops in the creation of pseudoscalar meson dressings of the
nucleon is easily observed. 
The self-energy considered below ($\Sigma_N$) is the total
contribution from those pion loops
which give rise to the leading non-analytic (LNA) and next-to-leading
non-analytic (NLNA) terms proportional to $F$ and $D$ in the
self-energy of the baryon, and
also the contributions that arise from the $\eta^{\prime}$ diagrams.
Explicitly we write the processes as $N \rightarrow N\pi \rightarrow N$, $N
\rightarrow \Delta\pi \rightarrow N$, $N \rightarrow N\eta^{\prime}
\rightarrow N$ and $N \rightarrow \Delta\eta^{\prime} \rightarrow N$. 
In the
limit of full QCD these $\eta^{\prime}$ contributions can be neglected 
because the $\eta^{\prime}$ is heavy and therefore decouples from the
low energy EFT.
The appearance of the unusual 
term $N \rightarrow \Delta\eta^{\prime} \rightarrow N$ is a consequence 
of the fact that intermediate states in the partially-quenched theory are 
not guaranteed to be physical. In particular, the ``$\Delta$'' here denotes 
a state of spin-3/2 (but not isospin-3/2) which is degenerate with the 
corresponding $\Delta$ state because the hyperfine gluon interaction 
depends only on the spin of the quark pairs.

For pQQCD in the heavy baryon limit the nucleon self-energy may be
expressed as:
\bea\label{eq:self_heavy_baryon}
\Sigma_{N} &=& \sigma_{NN}^{\pi} + \sigma_{NN}^{\eta^{\prime}} +
\sigma_{N\Delta}^{\pi} + \sigma_{N\Delta}^{\eta^{\prime}} 
\eea
Explicitly we have:
\bea
\label{eq:self_terms}
\sigma_{NN}^{\pi} &=& - \frac{3 (F + D)^{2}}{32 \pi f_{\pi}^{2}} 
\biggl( I(M_{PS}^{deg}, 0)\nonumber \\ & & \hspace{25mm} + \alpha 
\bigl(I(M_{PS}^{non-deg}, M_{N}^{non-deg} - M_{N}^{deg}) -
I(M_{PS}^{deg}, 0) \bigr) \biggr)
\nonumber \\ 
\nonumber \\
\sigma_{NN}^{\eta^{\prime}} &=& - \frac{(3F -D)^{2}}{32 \pi f_{\pi}^{2}} 
\biggl( \bigl((M_{PS}^{deg})^{2} - (M_{PS}^{unit})^{2} \bigr)I_{2}(M_{PS}^{deg})
\nonumber \\ & & \qquad \qquad \qquad
+\beta \bigl(I(M_{PS}^{non-deg}, M_{N}^{non-deg} - M_{N}^{deg}) - 
I(M_{PS}^{deg}, 0) \bigr)\biggr) \nonumber \\ \nonumber \\
\sigma_{N\Delta}^{\pi} &=& - \frac{1}{32 \pi f_{\pi}^{2}} \frac{8}{3}
\gamma^{2} \biggl( \frac{5}{8}I(M_{PS}^{deg}, M_{\Delta}^{deg} -
M_{N}^{deg}) \nonumber \\ & & \hspace{25mm}
+ \frac{3}{8}I(M_{PS}^{non-deg}, M_{\Delta}^{non-deg} -
M_{N}^{deg}) \biggr) \nonumber \\ \nonumber \\
\sigma_{N\Delta}^{\eta^{\prime}} &=& - \frac{1}{32 \pi f_{\pi}^{2}} \frac{1}{3}
\gamma^{2} \biggl(I(M_{PS}^{non-deg}, M_{\Delta}^{non-deg} - M_{N}^{deg}) \nonumber \\
& & \hspace{25mm} - I(M_{PS}^{deg}, M_{\Delta}^{deg} - M_{N}^{deg}) \biggr) 
\eea
As we will see (fig.\ref{fg:nucleon_se}), $\sigma_{NN}^\pi$ and 
$\sigma_{N\Delta}^\pi$ are typically negative, whereas
$\sigma_{NN}^{\eta^{\prime}}$ and 
$\sigma_{N\Delta}^{\eta^{\prime}}$ are typically around zero.
The parameters $\alpha,~\beta~\&~\gamma$ are derived from the standard
$SU(6)$ couplings \footnote{For a full discussion see \cite{su_six}.}.
Explicitly we take
\bea\label{eq:constants}
\alpha &=& \frac{\Gamma}{2(F+D)^{2}} \nonumber\\
\beta &=&  \frac{\Gamma}{2(3F-D)^2} \nonumber \\
\gamma &=& -2D \nonumber \\
\Gamma &=& \frac{1}{3} ( 3F + D )^2  +  3( D - F )^2
\eea
We use the constants $F=0.51$ and $D=0.76$, which are determined from fitting
semi-leptonic decays at tree level-- e.g., Ref.~\cite{bor}.

The integrals in eq.(\ref{eq:self_terms}) are defined as
\bea\label{eq:integrals}
I(M_{PS}, \delta M) &=& \frac{2}{\pi} \int_{0}^{\infty}
\frac{k^{4}u^{2}(k) dk}{\omega(\omega + \delta M)} \nonumber \\
I_{2}(M_{PS}) &=& \frac{2}{\pi} \int_{0}^{\infty} \frac{k^{4}u^{2}(k) dk}{\omega^{4}},
\eea
where we have used:
\bea
\omega(k)&=&\sqrt{k^{2} + M^{2}_{PS}}
\eea
Here $M_{PS}$ can be $M^{deg}_{PS}$ or $M^{non-deg}_{PS}$. We define 
this along
with values for $\delta M$ explicitly in the individual self-energy terms above.

We study both a standard dipole form factor, which takes the form
\bea\label{eq:dipole}
u(k)&=&\frac{\Lambda^{4}}{(\Lambda^{2}+k^{2})^2},
\eea
and a Gaussian form factor
\bea\label{eq:gaussian}
u(k)&=&\exp\Biggl({-\frac{k^2}{\Lambda^2}}\Biggr).
\eea

To account for finite volume artefacts, 
the self-energy equations are discretised so that only those momenta
allowed on the lattice appear:
\be
4 \pi \int_{0}^{\infty}k^2dk = \int d^3k \approx
\frac{1}{V}\left(\frac{2 \pi}{a}\right)^3 \sum_{k_x,k_y,k_z} \, ,
\label{eq:lat_int}
\ee
with
\be
k_{x,y,z} = \frac{2\pi(i,j,k)}{a N_{x,y,z}} \, .
\ee
with $i,j,k \in {\cal Z}$.
The purpose of the finite-range regulator is to regularise the theory
as $k_x$, $k_y$, $k_z$ tend to infinity. Indeed, once any one of
$k_x$, $k_y$ or $k_z$ is greater than $\sim 10\Lambda$ the
contribution to the integral is negligible and thereby ensuring
convergence of the summation.  Hence, we would like the highest
momentum in each direction to be just over $10\Lambda$.  For practical
calculation, we therefore use the following to calculate the maxima
and minima for i, j, k:
\bea
    (i,j,k)_{max} &=& ~~\left
    [\frac{10\Lambda~a}{2\pi}~N_{(x,y,z)}\right ] + 1, \nonumber \\
    (i,j,k)_{min} &=&  -\left [\frac{10\Lambda~a}{2\pi}~N_{(x,y,z)}
    \right ] - 1, \nonumber
\eea
where $[\ldots]$ denotes the integer part.

We now have the partially-quenched nucleon mass formula which we will
use in sec.~\ref{sec:fits} to fit the CP-PACS data.



\section{The CP-PACS Nucleon data}
\label{sec:cppacs}

\begin{figure}[*htbp] 
\begin{center} 
\includegraphics[angle=0, width=0.85\textwidth]{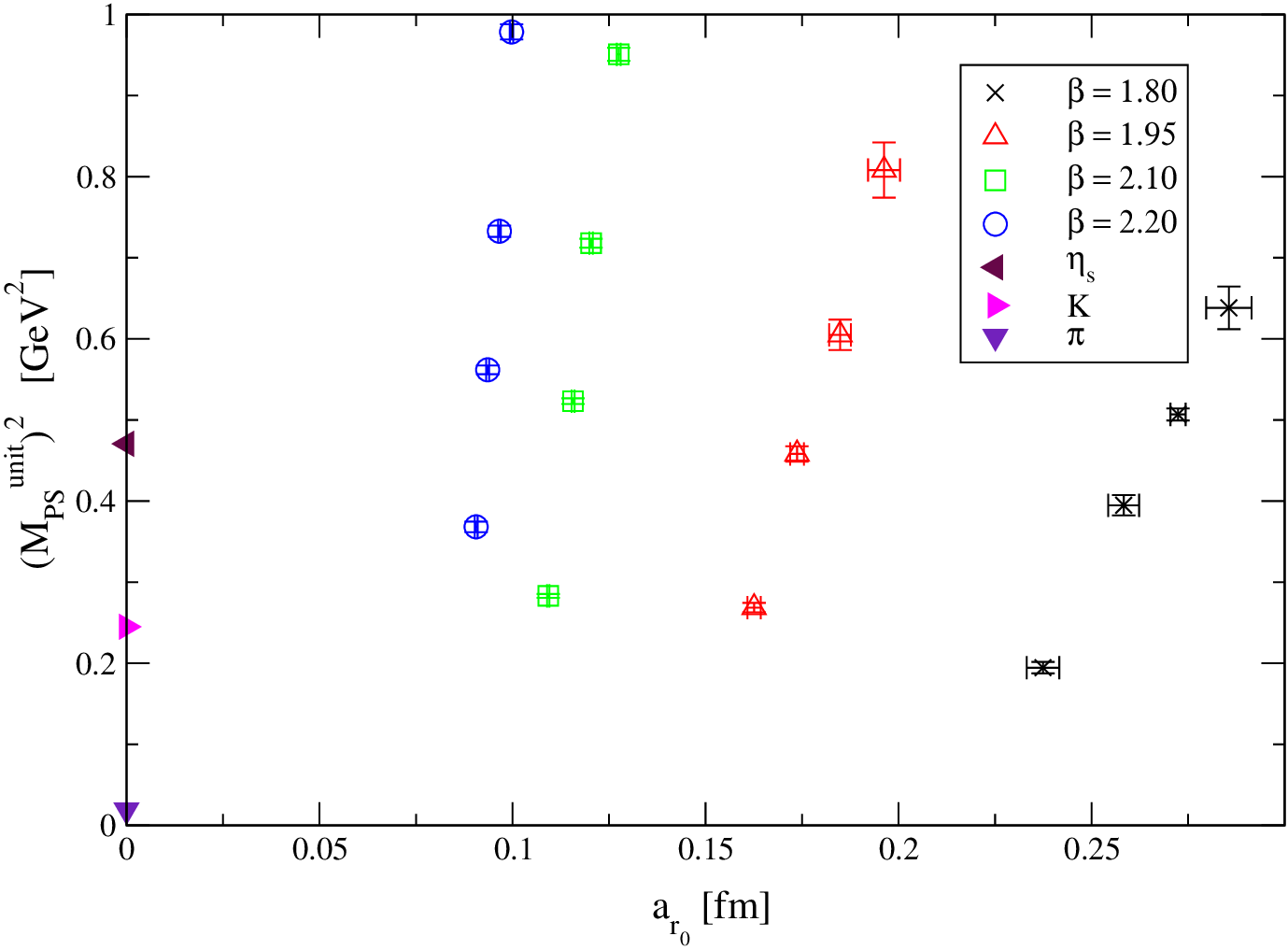}
\caption{The range of sea quark mass
$(M_{PS}^{unit})^2$ and lattice spacing, $a_{r_0}$, covered by the
\mbox{CP-PACS} data displayed in Table~\ref{tb:lat}.
$(M_{PS}^{unit})^2$ is the pseudoscalar meson mass squared
at the unitary point; i.e., where $\kval \equiv \ksea$.
The experimental points for the $\pi, K$ and ``$\eta_s$'' mesons
are also shown for reference.
\label{fg:a_r0_vs_mps2}} 
\end{center} 
\end{figure}

In Ref.~\cite{cppacs}, the CP-PACS collaboration published meson and
baryon spectrum data from dynamical simulations for mean-field
improved Wilson fermions with improved gluons at four different
$\beta$ values. For each value of $\beta$, ensembles were generated
for four values of $\ksea$ -- giving a total of 16
independent ensembles.  Table~\ref{tb:lat} summarises the lattice
parameters used and Figure \ref{fg:a_r0_vs_mps2} is a graphical
representation of the unitary pseudo-scalar masses plotted against the
lattice spacing $a_{r_{0}}$ and we note that
$(M_{PS}^{unit})^2$ is a direct measure of the sea quark mass.
For each of the sixteen ensembles there are five $\kval$ values considered
\cite{cppacs}. Thus there are a total of 80 $(M_N^{deg},M_{PS}^{deg})$
data points available for analysis.


\begin{table}
\begin{ruledtabular}
\begin{center}
\begin{tabular}{ccclll}
$\beta$ & $\kappa_{sea}$ & Volume & \multicolumn{1}{c}{$M_{PS}^{unit}/M_V^{unit}$} & \multicolumn{1}{c}{$a_{r_0}$ [fm]} & \multicolumn{1}{c}{$a_{\sigma}$ [fm]} \\
\hline
%
%
  1.80 & 0.1409 & $12^3 \times 24$ &  0.8067\err{ 9}{ 9} & 0.286\err{ 6}{ 6}  & 0.288\err{ 3}{ 3}   \\
  1.80 & 0.1430 & $12^3 \times 24$ &  0.7526\err{16}{15} & 0.272\err{ 2}{ 2}  & 0.280\err{ 4}{ 5}   \\
  1.80 & 0.1445 & $12^3 \times 24$ &  0.694\err{ 2}{ 2}  & 0.258\err{ 4}{ 4}  & 0.269\err{ 2}{ 3}   \\
  1.80 & 0.1464 & $12^3 \times 24$ &  0.547\err{ 4}{ 4}  & 0.237\err{ 4}{ 4}  & 0.248\err{ 2}{ 3}   \\
\hline
  1.95 & 0.1375 & $16^3 \times 32$ &  0.8045\err{11}{11} & 0.196\err{ 4}{ 4}  & 0.2044\err{10}{12}   \\
  1.95 & 0.1390 & $16^3 \times 32$ &  0.752\err{ 2}{ 2}  & 0.185\err{ 3}{ 3}  & 0.1934\err{14}{15}   \\
  1.95 & 0.1400 & $16^3 \times 32$ &  0.690\err{ 2}{ 2}  & 0.174\err{ 2}{ 2}  & 0.1812\err{12}{12}   \\
  1.95 & 0.1410 & $16^3 \times 32$ &  0.582\err{ 3}{ 3}  & 0.163\err{ 2}{ 2}  & 0.1699\err{13}{15}   \\
\hline
  2.10 & 0.1357 & $24^3 \times 48$ &  0.806\err{ 2}{ 2}  & 0.1275\err{ 5}{ 5} & 0.1342\err{ 8}{ 8}   \\
  2.10 & 0.1367 & $24^3 \times 48$ &  0.755\err{ 2}{ 2}  & 0.1203\err{ 4}{ 5} & 0.1254\err{ 8}{ 8}   \\
  2.10 & 0.1374 & $24^3 \times 48$ &  0.691\err{ 3}{ 3}  & 0.1157\err{ 4}{ 4} & 0.1203\err{ 6}{ 6}   \\
  2.10 & 0.1382 & $24^3 \times 48$ &  0.576\err{ 3}{ 4}  & 0.1093\err{ 3}{ 3} & 0.1129\err{ 4}{ 5}   \\
\hline
  2.20 & 0.1351 & $24^3 \times 48$ &  0.799\err{ 3}{ 3}  & 0.0997\err{ 4}{ 5} & 0.10503\err{15}{15}  \\
  2.20 & 0.1358 & $24^3 \times 48$ &  0.753\err{ 4}{ 4}  & 0.0966\err{ 4}{ 4} & 0.1013\err{ 3}{ 2}   \\
  2.20 & 0.1363 & $24^3 \times 48$ &  0.705\err{ 6}{ 6}  & 0.0936\err{ 4}{ 4} & 0.0978\err{ 3}{ 3}   \\
  2.20 & 0.1368 & $24^3 \times 48$ &  0.632\err{ 8}{ 8}  & 0.0906\err{ 4}{ 4} & 0.0949\err{ 2}{ 2}   \\
\end{tabular} 
\end{center} 
\end{ruledtabular}
\caption{\small The lattice parameters of the CP-PACS simulation used
in this data analysis, taken from Ref.~\cite{cppacs}.  The superscript
{\em unit} refers to the unitary data (i.e., where $\kval^1 \equiv
\kval^2 \equiv \ksea$).  Note that the errors reported in this table
are obtained with our bootstrap ensembles (see text).
\label{tb:lat}}
\end{table}



%
%


The analysis in this paper shares many common features with our analysis
of the vector meson data in Ref.~\cite{rhopapers}. In particular, we
consider two methods of setting the scale: the string tension
$(\sigma)$, and the hadronic scale $(r_{0})$.  We generate 1000
bootstrap clusters for all hadronic masses from a Gaussian
distribution with a central value equal to the values published in
\cite{cppacs} (tables XXI, XXII and XXIII \footnote{We take the values
  of $m_{\Sigma}$ in table XXIII to be the mass values for the
  non-degenerate nucleon. We can do this since the interpolation
  operators for $N$ and $\Sigma$ have the same quantum numbers}) and a
FWHM equal to the published error.  We use totally uncorrelated data
throughout, which we argue in Ref.~\cite{rhopapers} leads to our statistical
errors being overestimates.  The values $r_{0} = 0.49$ fm and
$\sqrt{\sigma} = 440$ MeV are used.

Since the action used in \cite{cppacs} is mean-field, rather than
non-perturbatively improved, it will have some residual lattice
systematic errors of ${\cal O}(a)$.  We therefore fit the data
assuming both ${\cal O}(a)$ and ${\cal O}(a^2)$ effects, which we
investigate in Sec.~\ref{sec:fits}.

The physical volume for the $\beta=1.80, 1.95$ and $2.10$ ensembles is
$La \approx 2.5$ fm, and the $\beta=2.20$ ensemble has a
slightly smaller physical volume.  The associated finite volume
effects are incorporated through evaluating the chiral loops by
explicitly summing the discrete pion momenta allowed on the lattice
as described in eq.(\ref{eq:lat_int}).



\section{Fitting Analysis}
\label{sec:fits}

\subsection{Summary of analysis techniques}
\label{sec:nucleon_fits_summary}

The philosophy behind our fitting method remains the same as for our
investigation of the meson spectrum~\cite{rhopapers}, i.e. we work in
physical units when performing our extrapolations. We do this so
that data from different ensembles can be combined; something which cannot
be done for the dimensionless data since they correspond to differing
lattice spacings.  In addition, we expect that 
this approach will benefit from some
cancellation of the systematic (and statistical) errors.

The Adelaide approach to chiral fits describes the variation of hadron
mass with quark mass by a combination of the self-energy term (in this
case, $\Sigma_N$) with ``constituent quark'' terms
(i.e. polynomials in the {\em valence} quark mass). The former
accurately describe the chiral behaviour and become negligible as the
quark mass becomes heavy.  Thus we have
\bea\label{eq:nucleon_adel_fit_form}
M_{N} - \Sigma_{N} &=& a_{0} + a_{2} (M_{PS}^{deg})^2 + a_{4} (M_{PS}^{deg})^4 +
a_{6} (M_{PS}^{deg})^6 \,.
\eea
In the complete EFT for the partially-quenched theory, there are
contributions to this polynomial expansion which measure the
displacement from the unitarity point (where $m_{val}=m_{sea}$).
This freedom could be
incorporated by extending the terms polynomial in the quark mass by
\begin{eqnarray}
a_2 (M_{PS}^{deg})^2 &\to& a_2 (M_{PS}^{deg})^2 + a_2' \delta_{vs}^2\,, \\
\nonumber
a_4 (M_{PS}^{deg})^4 &\to& a_4 (M_{PS}^{deg})^4 + a_4' (M_{PS}^{deg})^2 \delta_{vs}^2 + a_4'' \delta_{vs}^4\,,
\end{eqnarray}
and similarly for the third-order term. Here, the notation
$\delta_{vs}^2=(M_{PS}^{deg})^2-(M_{PS}^{unit})^2$ is used to describe
the mass splitting between the sea and valence quarks. In line with
our earlier work, Ref.~\cite{rhopapers}, we make a model assumption by
ignoring the terms proportional to $\delta_{vs}^2$.  We have checked
this assumption by confirming numerically that there is no variation
of the $a_2$ coefficient (obtained with the ``individual ensemble
fit'' approach --- see next subsection) with $M_{PS}^{unit}$. 
We found a similar situation in our earlier work, Ref.~\cite{rhopapers}.
 In any
case, any dependencies below the level of our
statistics on $\delta_{vs}^2$ are implicitly contained within the
chiral self-energies, $\Sigma_N$. Thereby, we use the lattice data to
select a preferential regularisation scale (and perhaps regulator)
which efficiently interpolates between the partially-quenched and
unitary points.
This method has proven very successful in connecting quenched and
dynamical QCD results over a range of observables, see
Refs.~\cite{Young:2002cj,Young:2004tb,Leinweber:2004tc,Leinweber:2006ug}
for example.
In effect, the 6 new parameters (at this order)
characterising the non-unitarity dependence are modelled by a single
parameter $\Lambda$.

Again, following our vector-meson analysis \cite{rhopapers}, we
contrast this chirally-motivated approach with a naive polynomial
fitting function,
\bea\label{eq:nucleon_fit_form}
M_{N} &=& a_{0} + a_{2} (M_{PS}^{deg})^2 + a_{4} (M_{PS}^{deg})^4 + a_{6}
(M_{PS}^{deg})^6 \,.
\eea
We divide these fits into two categories, ``cubic'' and ``quadratic''
depending on whether or not the $(M_{PS}^{deg})^6$ is included.

\begin{figure}[*htbp]
\begin{center}
\includegraphics[angle=0, width=0.85\textwidth]
{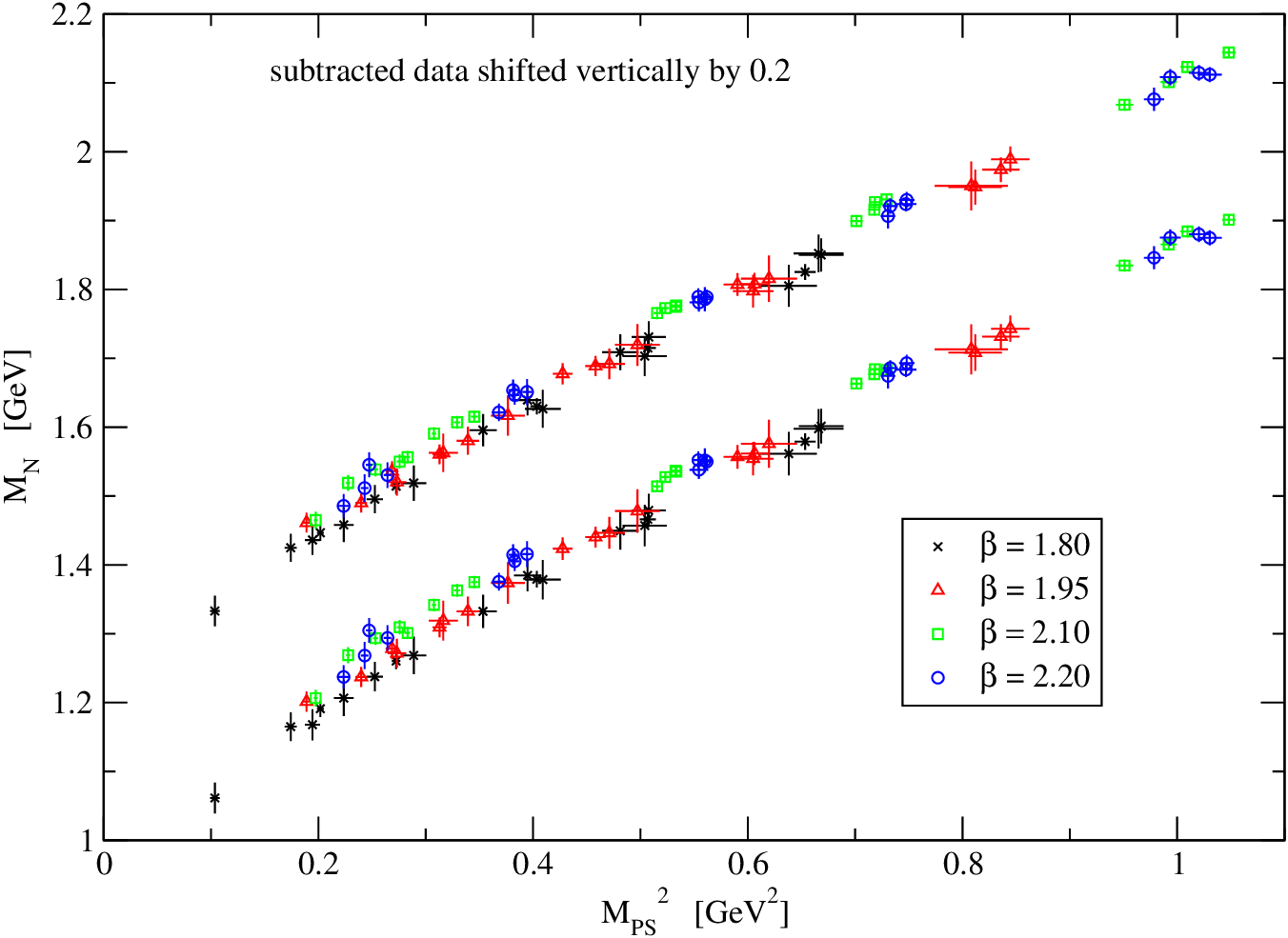}
\caption
{A plot of the degenerate \mbox{CP-PACS} nucleon data set. Here the
  scale is set using $r_{0}$. Both the raw data, $M_N$, and the
  subtracted data, $M_N - \Sigma_N$, are shown, with the latter shifted
  vertically by 0.2 GeV for clarity. A dipole form factor with $\Lambda =
  600$ MeV was chosen in the latter case.
\label{fg:global_nucleon_raw_and_subtracted_r0}}
\end{center}
\end{figure}

In fig.~\ref{fg:global_nucleon_raw_and_subtracted_r0} we plot the
dimensionful nucleon data, $M_N$, and the subtracted nucleon data,
$M_N - \Sigma_N$. In the latter, we use a representative value of
$\Lambda=600$ MeV with the dipole form factor and the scale set by
$r_0$.


\begin{figure}[*htbp] 
\begin{center} 
\includegraphics[angle=0, width=0.85\textwidth]{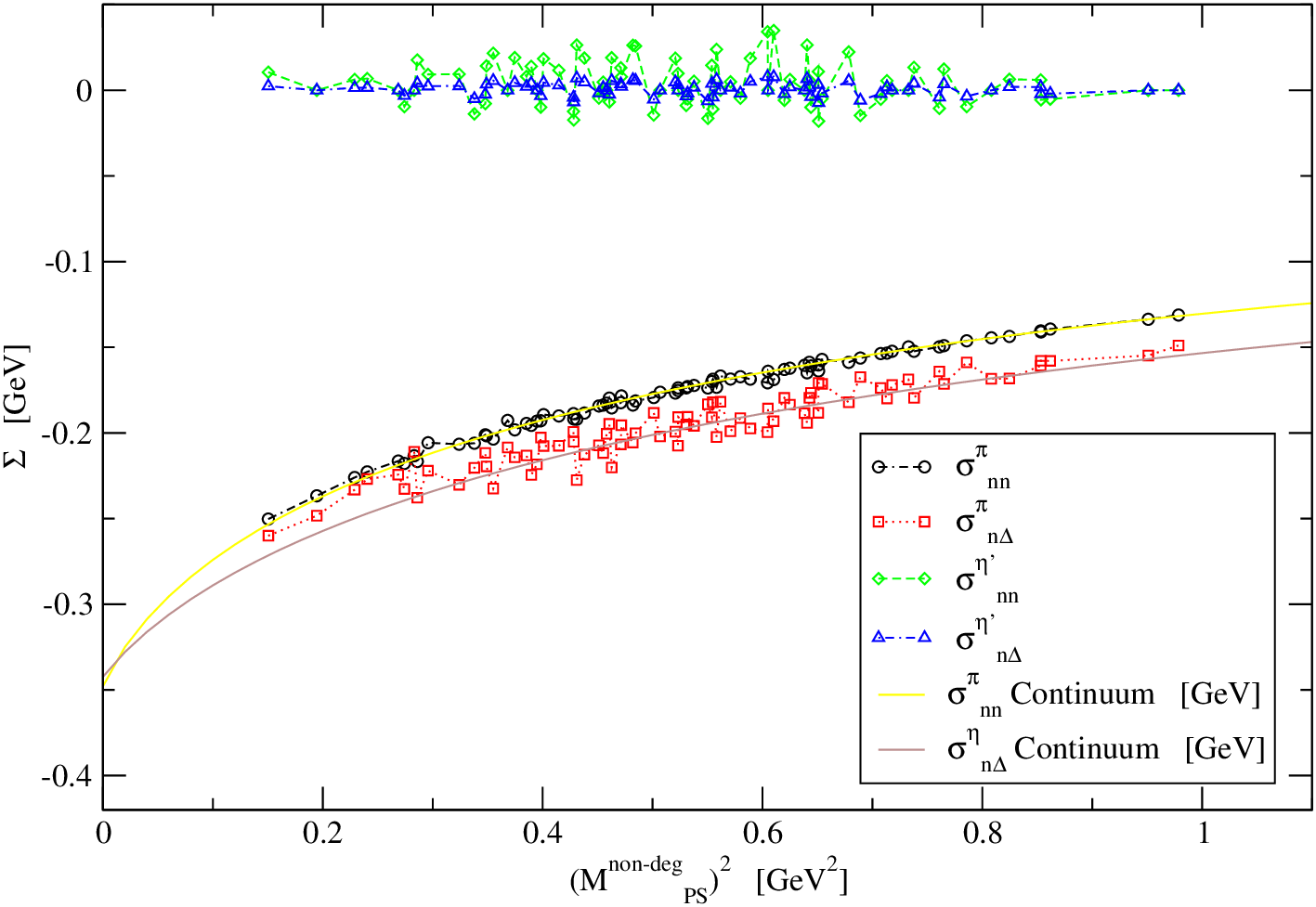}
\caption[A plot of the self-energy contributions for the nucleon versus
$(M_{PS}^{non-deg})^{2}$.]{Here we plot the self-energy contributions 
Eqs.~\ref{eq:self_terms}) versus $(M_{PS}^{non-deg})^{2}$ for 
the entire degenerate
data set (dashed lines are a guide for the eye only). We use the dipole form
factor and  choose an arbitrary value for the Lambda parameter, $\Lambda = 1$
[GeV]. We also include continuum, physical data (the solid curves) for the pion
processes (the $\eta$ case vanishes in the physical limit).
\label{fg:nucleon_se}} 
\end{center} 
\end{figure}

We plot, in fig.~\ref{fg:nucleon_se}, the terms which make up
$\Sigma_N$ (see eq.(\ref{eq:self_heavy_baryon})) for each of the 80
data points under consideration, in order to get a feel for their
relative size. We also show the continuum, physical values for these
terms, noting that in this case, the terms involving $\eta^{\prime}$
vanish (as required). The curvature in the terms which make up
$\Sigma_N$ seen in fig.~\ref{fg:nucleon_se} as $M_{PS}\rightarrow 0$
matches, by design, the LNA and NLNA chiral contributions (see
sec.\ref{sec:se}). The polynomial chiral fits obviously do not
reproduce this chiral behaviour. It is this curvature which means that
the Adelaide approach more accurately reproduces the experimental
value of the nucleon mass than the naive polynomial approach,
predicting a nucleon mass some 60MeV lower than the polynomial
approach (see later).


%
%


In the next subsection we individually fit
Eqs.~(\ref{eq:nucleon_adel_fit_form} \& \ref{eq:nucleon_fit_form}) to
the sixteen ensembles (Table~\ref{tb:lat}). Following this we perform
a single global fit of the entire data set using
Eqs.~(\ref{eq:nucleon_adel_fit_form} \& \ref{eq:nucleon_fit_form})
modified appropriately by ${\cal O}(a)$-style corrections.



\subsection{Individual ensemble fits}
\label{sec:nucleon_individual}

In this section we treat the sixteen ensembles separately by fitting
Eqs.(\ref{eq:nucleon_adel_fit_form} \& \ref{eq:nucleon_fit_form}) to
the five degenerate data points $(M^{deg}_{N}, M^{deg}_{PS})$ from
each of the sixteen ensembles.  In the Adelaide case, we have used our
preferred choices of the dipole form factor with the scale taken from
$r_{0}$.  The other possibilities (i.e., using the Gaussian form factor
and taking the scale from the string tension) are discussed in
Sec.~\ref{sec:nucleon_global}.  We have chosen $\Lambda=600$ MeV for
these individual fits which is very close to what turns out to be our
preferred value in Sec.~\ref{sec:nucleon_global}.  (Section
\ref{sec:nucleon_global} discusses in detail the variation of nucleon
mass with $\Lambda$.)  The fits considered in this section are
quadratic in the chiral expansion (i.e., we set $a_{6} = 0$ in
Eqs.~(\ref{eq:nucleon_adel_fit_form} \& \ref{eq:nucleon_fit_form}))
since cubic fits for the individual fits have 100\% error in the
$a_{4}$ and $a_{6}$ coefficients.



\begin{table}
\begin{ruledtabular}
\begin{center}
\begin{tabular}{cccllccccc}
\hline
&&&&&&&&&\\
 & $\beta$ & $\kappa_{sea}$ & \multicolumn{1}{c}{$a^{naive}_{0}$} & \multicolumn{1}{c}{$a^{adel}_{0}$} & 
$a^{naive}_{2}$ & $a^{adel}_{2}$ & $a^{naive}_{4}$ & $a^{adel}_{4}$ & \\
 & & & [GeV] & [GeV] & [GeV$^{-1}$] & [GeV$^{-1}$] & [GeV$^{-3}$] & [GeV$^{-3}$] \\
&&&&&&&&&\\
\hline
&&&&&&&&&\\
%
%
  & 1.80 & 0.1409 &  0.97\err{ 4}{ 4} &  1.03\err{ 4}{ 4} & 1.11\err{12}{14} & 
 1.09\err{13}{14} & -0.30\err{15}{13} & -0.29\err{15}{14} & \\
  & 1.80 & 0.1430 &  0.98\err{ 3}{ 2} &  1.04\err{ 3}{ 2} & 1.11\err{11}{12} & 
 1.08\err{11}{12} & -0.29\err{13}{12} & -0.29\err{13}{13} & \\
  & 1.80 & 0.1445 &  0.96\err{ 3}{ 3} &  1.03\err{ 3}{ 3} & 1.21\err{12}{11} & 
 1.18\err{13}{12} & -0.37\err{13}{14} & -0.37\err{13}{14} & \\
  & 1.80 & 0.1464 &  0.93\err{ 3}{ 3} &  1.01\err{ 3}{ 3} & 1.27\err{12}{10} & 
 1.23\err{12}{11} & -0.42\err{12}{14} & -0.41\err{12}{14} & \\
&&&&&&&&&\\
  & 1.95 & 0.1375 &  1.00\err{ 4}{ 3} &  1.05\err{ 4}{ 3} & 1.08\err{10}{12} & 
 1.07\err{11}{12} & -0.25\err{10}{ 8} & -0.25\err{10}{ 9} & \\
  & 1.95 & 0.1390 &  1.00\err{ 3}{ 2} &  1.06\err{ 3}{ 2} & 1.04\err{ 8}{ 8} & 
 1.02\err{ 8}{ 8} & -0.21\err{ 7}{ 7} & -0.20\err{ 7}{ 7} & \\
  & 1.95 & 0.1400 &  0.99\err{ 2}{ 2} &  1.05\err{ 2}{ 2} & 1.11\err{ 7}{ 7} & 
 1.08\err{ 7}{ 7} & -0.26\err{ 6}{ 6} & -0.25\err{ 6}{ 6} & \\
  & 1.95 & 0.1410 &  1.01\err{ 2}{ 2} &  1.07\err{ 2}{ 2} & 1.08\err{ 7}{ 6} & 
 1.05\err{ 7}{ 6} & -0.24\err{ 6}{ 6} & -0.23\err{ 6}{ 7} & \\
&&&&&&&&&\\
  & 2.10 & 0.1357 &  1.04\err{ 2}{ 2} &  1.08\err{ 2}{ 2} & 1.06\err{ 7}{ 7} & 
 1.05\err{ 7}{ 7} & -0.23\err{ 5}{ 5} & -0.23\err{ 5}{ 5} & \\
  & 2.10 & 0.1367 &  1.05\err{ 2}{ 2} &  1.10\err{ 2}{ 2} & 1.01\err{ 7}{ 7} & 
 0.99\err{ 7}{ 7} & -0.19\err{ 5}{ 5} & -0.19\err{ 5}{ 5} & \\
  & 2.10 & 0.1374 &  1.04\err{ 2}{ 2} &  1.10\err{ 2}{ 2} & 1.03\err{ 7}{ 7} & 
 1.01\err{ 7}{ 7} & -0.19\err{ 5}{ 5} & -0.19\err{ 5}{ 5} & \\
  & 2.10 & 0.1382 &  1.00\err{ 2}{ 2} &  1.06\err{ 2}{ 2} & 1.13\err{ 6}{ 6} & 
 1.10\err{ 6}{ 6} & -0.25\err{ 4}{ 4} & -0.25\err{ 4}{ 4} & \\
&&&&&&&&&\\  
  & 2.20 & 0.1351 & 1.04\err{ 5}{ 5} & 1.08\err{ 5}{ 5} &
  1.0\err{ 2}{ 2} & 
 1.0\err{ 2}{ 2} & -0.21\err{12}{15} & -0.21\err{13}{15} & \\
  & 2.20 & 0.1358 & 1.10\err{ 4}{ 4} & 1.14\err{ 4}{ 4} & 0.87\err{12}{13} & 
 0.86\err{12}{14} & -0.09\err{10}{ 9} & 
 -0.08\err{10}{ 9} & \\
  & 2.20 & 0.1363 & 1.04\err{ 4}{ 4} & 1.08\err{ 4}{ 4} & 1.03\err{13}{11} & 
 1.01\err{13}{12} & -0.20\err{ 8}{ 9} & -0.19\err{ 8}{ 9} & \\
  & 2.20 & 0.1368 & 1.01\err{ 4}{ 3} & 1.06\err{ 4}{ 3} & 1.08\err{11}{11} & 
 1.05\err{11}{11} & -0.23\err{ 8}{ 8} & -0.22\err{ 8}{ 8} & \\
&&&&&&&&&\\
\hline 
\end{tabular} 
\end{center} 
\end{ruledtabular}
\caption[The coefficients obtained from fitting $M_{N}$ data against $M_{PS}^2$.]
{The coefficients obtained from fitting $M_{N}$ data against $M_{PS}^2$.
We list results for both the naive and Adelaide fits (
Eqs.~(\ref{eq:nucleon_fit_form}) \& (\ref{eq:nucleon_adel_fit_form}), 
respectively) for
each of the sixteen ensembles listed in Table~\ref{tb:lat}.
A dipole form factor was employed for the Adelaide fits using $\Lambda = 600$
[MeV] and the scale was set from $r_0$.
\label{tb:nucleon_individual}}
\end{table}
%

Table~\ref{tb:nucleon_individual} lists the coefficients for both the
Adelaide fits and also the naive fits using this approach. As expected
the leading Adelaide coefficient is always greater than the
corresponding coefficient from the naive fits $(a^{adel}_{0} >
a^{naive}_{0})$.  In virtually all cases the $a_{2}$ coefficient is
smaller for the Adelaide fits $(a^{adel}_{2} < a^{naive}_{2})$. The
$a_{4}$ coefficients are approximately the same for both fits
$(a^{adel}_{4} \sim a^{naive}_{4})$, but the error in this coefficient
is very large, typically 50\%. We see only that the $a_4$ coefficient
is zero within errors in only one ensemble, indicating its presence is
needed.

\begin{figure}[*htbp]
\begin{center}
\includegraphics[angle=0, width=0.85\textwidth]{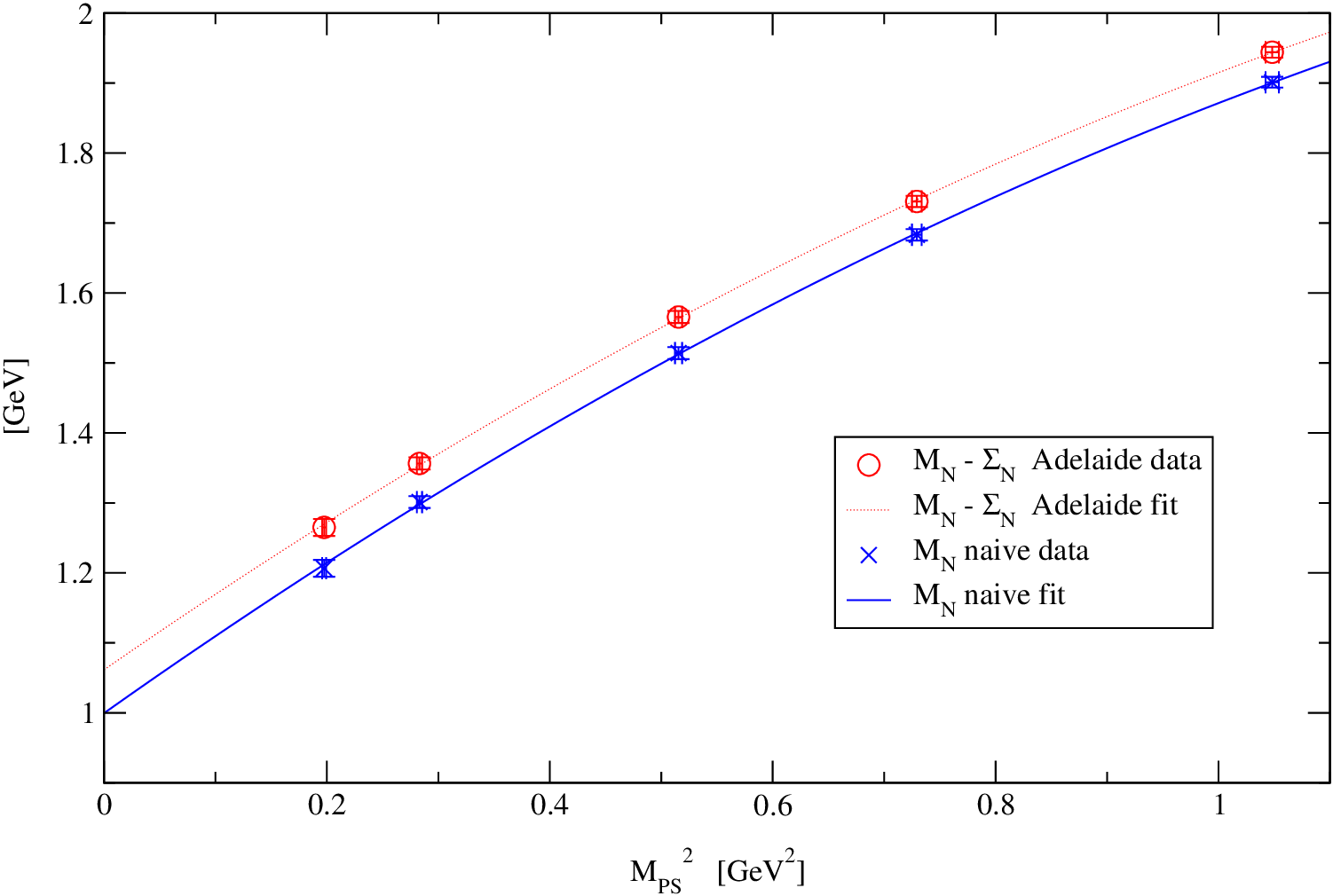}
\caption[A plot of $M_{N}$ versus $M_{PS}^{2}$ for the second lightest
  ensemble.  Included are the results of the quadratic naive and the
  quadratic Adelaide fits.]  {A plot of $M_{N}$ versus $M_{PS}^{2}$
  for the ensemble $(\beta,\ksea) = (2.10,0.1382)$. Included are the
  results of the quadratic naive, Eq.~(\ref{eq:nucleon_fit_form}) and
  the quadratic Adelaide, eq.(\ref{eq:nucleon_adel_fit_form})
  fits. The scale is set from $r_{0}$, we use a dipole form factor and
  our preferred value for $\Lambda$ ($\Lambda = 600$ [MeV]).
\label{fg:second_lightest_nucleon}}
\end{center}
\end{figure}

{}Figure~\ref{fg:second_lightest_nucleon} is a representative example of
one of these fits. It corresponds to the $(\beta, \ksea) = (2.10,0.1382)$
ensemble, which is one of closest to the physical point (
{}fig.~\ref{fg:a_r0_vs_mps2}).


%
%



%
\begin{figure}[*htbp]
\begin{center}
\includegraphics[angle=0, width=0.85\textwidth]{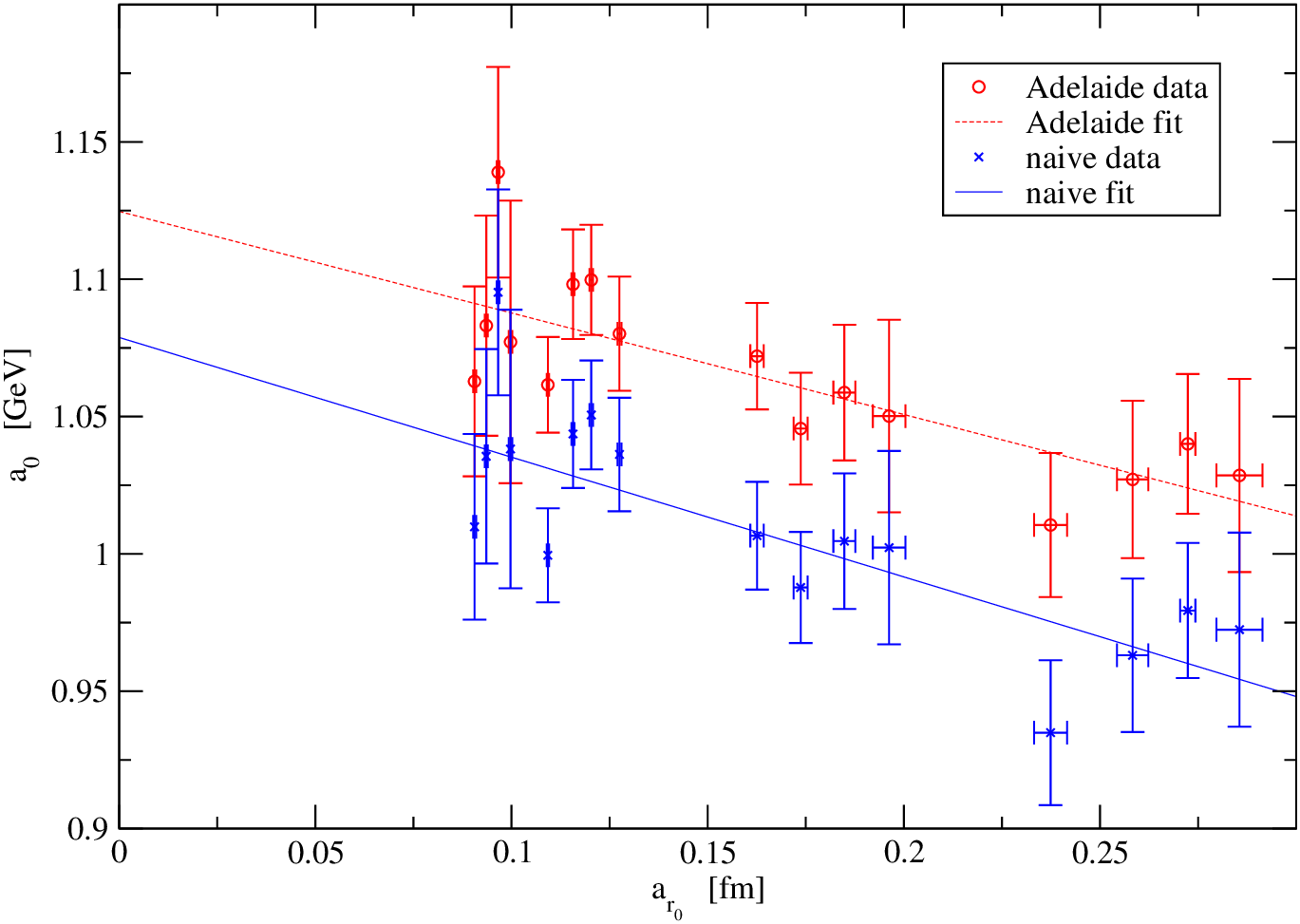}
\caption
{A continuum extrapolation of the $a_0$ coefficient obtained from both
the Adelaide and naive fits Eq.~(\ref{eq:a02_cont_nucleon}).
\label{fg:a0_vs_a_r0_both_nucleon}}
\end{center}
\end{figure}
%



%
\begin{figure}[*htbp]
\begin{center}
\includegraphics[angle=0, width=0.85\textwidth]{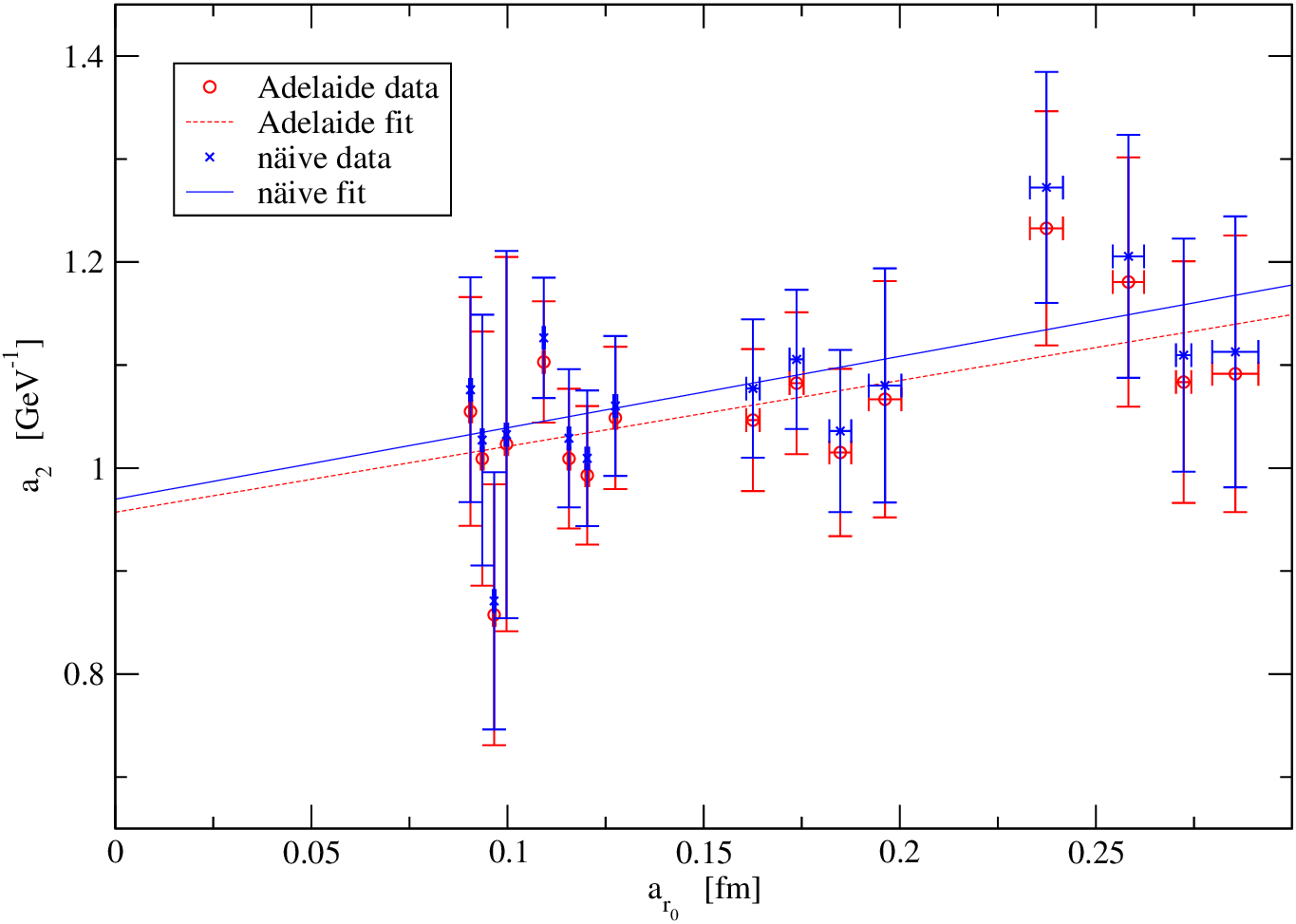}
\caption
{A continuum extrapolation of the $a_2$ coefficient obtained from both
the Adelaide and naive fits Eq.~(\ref{eq:a02_cont_nucleon}).
\label{fg:a2_vs_a_r0_both_nucleon}}
\end{center}
\end{figure}
%


With a view to performing a combined, global fit to all sixteen
ensembles, we note from Table~\ref{tb:nucleon_individual} that there
appears to be no discernible trend with the sea quark mass for any of
the chiral coefficients. However, there does appear to be a lattice
spacing effect. Figures~(\ref{fg:a0_vs_a_r0_both_nucleon} \&
\ref{fg:a2_vs_a_r0_both_nucleon}) plot $a_{0,2}$ against the lattice
spacing from $r_0$, $a_{r_{0}}$.  These plots motivate the following
continuum extrapolation
\be\label{eq:a02_cont_nucleon}
a_{0,2} = a_{0,2}^{cont} + X^{individual}_{0,2} \;a_{r_0}.
\ee
(Note that in sec.~\ref{sec:nucleon_global} we investigate both ${\cal
  O}(a)$ and ${\cal O}(a^2)$ corrections to the chiral coefficients.)
The results of the fits corresponding to
Eq.~(\ref{eq:a02_cont_nucleon}) are listed in 
Table~\ref{tb:a02_cont_nucleon}.


\begin{table}
\begin{center}
\begin{tabular}{l|ccc|ccc}
\hline
&&&&&&\\
  & $a_{0}^{cont.}$ & $X^{individual}_{0}$ & $\chi^{2}_{0}/d.o.f.$ &
    $a_{2}^{cont.}$ & $X^{individual}_{2}$ &  $\chi^{2}_{2}/d.o.f.$ \\
  & [GeV]           & [GeV/fm]             &                       &
    [GeV$^{-1}$]    & [GeV$^{-1}$/fm] & \\
&&&&&&\\
\hline
&&&&&&\\
Naive-fit  & 1.08\err{ 2}{ 2}  & -0.44\err{10}{11} & 13 / 14 & 0.97\err{ 7}{ 6}
& 0.7\err{ 4}{ 4}   & 7 / 14  \\
&&&&&&\\
Adelaide-fit & 1.12\err{ 2}{ 2}  & -0.37\err{10}{11} & 8 / 14 & 0.96\err{ 7}{
6} & 0.6\err{ 4}{ 4} & 6 / 14 \\
&&&&&&\\
\hline
\end{tabular}
\end{center}
\caption{The coefficients obtained from the continuum extrapolation of
both the naive and Adelaide $a_{0,2}$ values from 
Table~\ref{tb:nucleon_individual} using Eq.~(\ref{eq:a02_cont_nucleon}).
\label{tb:a02_cont_nucleon}}
\end{table}


We note that the errors in the $X^{individual}_0$ coefficients are
around 25\%, whereas they are more than 50\% for $X^{individual}_2$.




\subsection{Global fits}
\label{sec:nucleon_global}

We now analyse the complete set of 80 $(M_{PS}, M_N)$ data points by
globally fitting the sixteen ensembles of Table~\ref{tb:lat}.  As in
Ref.~\cite{rhopapers}, the idea is that this will produce a highly
constrained fit and allow us to determine the higher order
coefficients in the chiral expansion in
Eqs.(\ref{eq:nucleon_adel_fit_form} \& \ref{eq:nucleon_fit_form}), the
Adelaide scale parameter, $\Lambda$, and the preferred form factor.


In order to combine data from different ensembles into a single fit, we
use the experience gained in sec.~\ref{sec:nucleon_individual} and in
Refs.~\cite{rhopapers}. This tells us that the data's lattice spacing
artifacts, which are sizable enough for us to discern, lie in the
leading coefficient, $a_0$.  We have checked this conclusion by
studying combinations of ${\cal O}(a)$ and ${\cal O}(a^2)$ terms
\footnote{We choose ${\cal O}(a)$ and ${\cal O}(a^{2})$ corrections
  because the lattice action is tree-level improved and so should
  contain ${\cal O}(a^{2})$ errors together with some ${\cal O}(a)$
  errors.}  in the $a_2$ and higher coefficients, but have found that
these fits are unstable. Therefore our global fitting functions for
the Adelaide case is a modified version of
Eq.~(\ref{eq:nucleon_adel_fit_form}):
\bea\label{eq:nucleon_adel_fit_global}
\hspace{-12mm}
M_{N} - \Sigma_{N} &=& (a_{0} + X_{n} a^{n}) + a_{2} (M_{PS}^{deg})^2 + a_{4}
(M_{PS}^{deg})^4 + a_{6} (M_{PS}^{deg})^6
\eea
and the global fit function corresponding to the naive case
Eq.~(\ref{eq:nucleon_fit_form}) is
\bea\label{eq:nucleon_fit_global}
M_{N} &=& (a_{0}  + X_{n} a^{n}) + a_{2} (M_{PS}^{deg})^2 + a_{4}
(M_{PS}^{deg})^4 + a_{6} (M_{PS}^{deg})^6
\eea

We consider both ``quadratic'' and ``cubic'' chiral fits and also
consider both ${\cal O}(a)$ and ${\cal O}(a^2)$ lattice spacing
effects in the $a_{0}$ coefficient.  Hence, for the global fit analysis
the maximum number of fit parameters in any one fitting function is
five. (We discuss the $\Lambda$ parameter in the Adelaide case later.)
Since our data set contains 80 points, the fit will hopefully provide
highly constrained fit parameters compared to those from the
individual fitting method (see Table~\ref{tb:nucleon_individual}). The
scale was set using both the string tension and Sommer scale, $r_0$.
Finally, for the Adelaide method we study both the dipole,
Eq.~(\ref{eq:dipole}), and Gaussian, Eq.~(\ref{eq:gaussian}) form
factors.

The different choices of fitting procedure are summarised in 
Table~\ref{tb:fit_types_nucleon}. In this table, the entries in
each column represent a separate possibility, 
making making a total of $2^4$ Adelaide
and $2^3$ naive fitting types.


\begin{table}
\begin{ruledtabular}
\begin{center}
\begin{tabular}{cccc}
Fit                & Chiral          & ${\cal O}(a^n)$ & Lattice \\ 
Approach           & Extrapolation   & term in $a_0$   & Spacing from: \\
\hline
Adelaide - Dipole  &                 &                 & \\
                   & Cubic           & ${\cal O}(a)$   & $r_0$ \\
Adelaide - Gaussian&                 &                 & \\
                   & Quadratic       & ${\cal O}(a^2)$ & $\sigma$ \\
Naive              &                 &                 & \\
\end{tabular}
\end{center}
\end{ruledtabular}
\caption{The different fit types used in the global analysis. 
\label{tb:fit_types_nucleon}}
\end{table} 

\begin{figure}[*htbp]
\begin{center}
\includegraphics[angle=0, width=0.85\textwidth]
{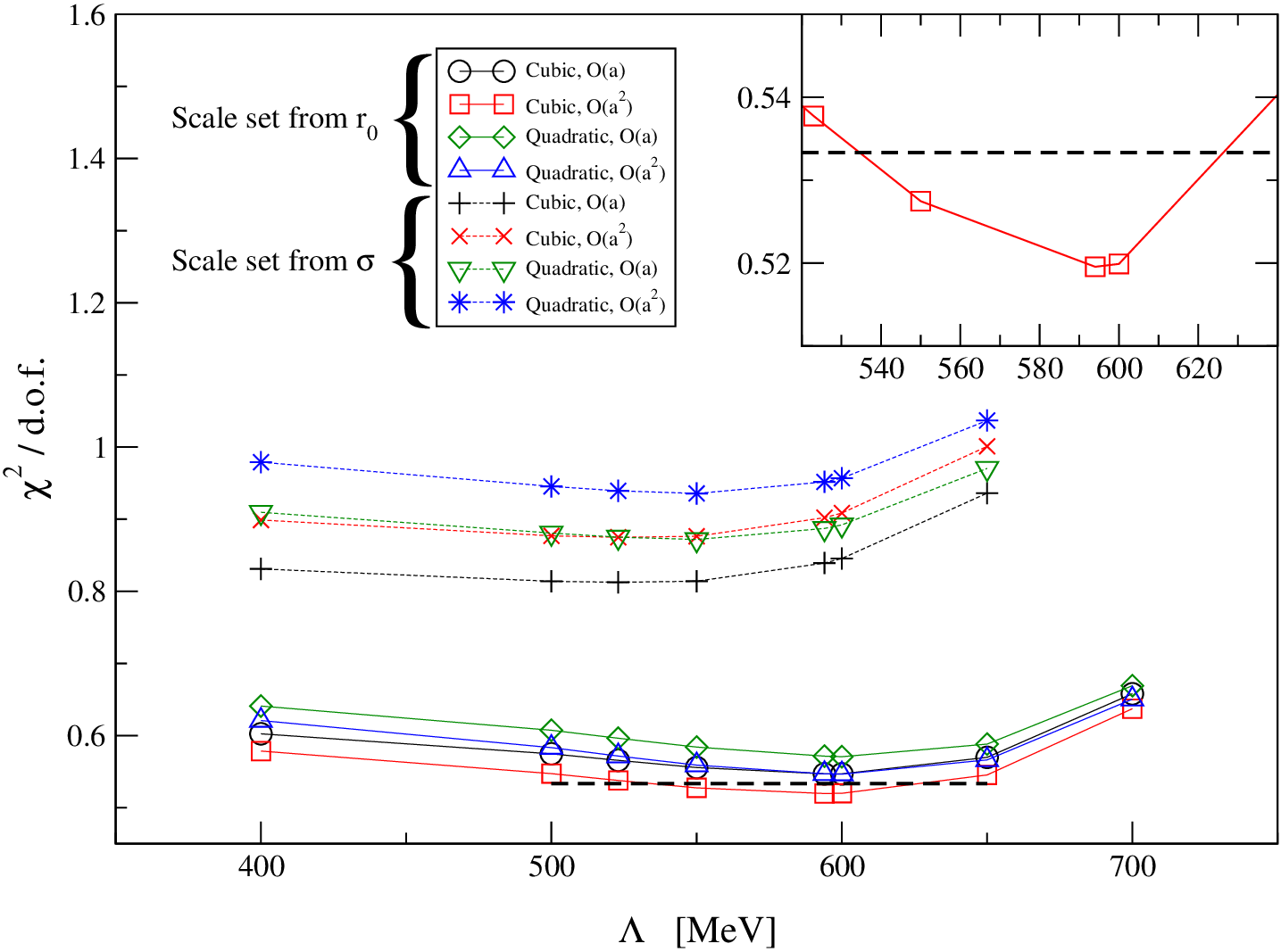}
\caption[A plot of $\chi^2 / d.o.f$ against $\Lambda$ using a dipole
  form factor for the nucleon case.]  {A plot of $\chi^2 / d.o.f$
  against $\Lambda$ for the dipole form factor.  The dashed horizontal
  line represents increasing $\chi^2$ from its minimum value by unity
  for the $r_0$ data (i.e. it represents one standard deviation). The
  intercept of this dashed line with the $\chi^2$ curves (at $\Lambda
  = $535 and 626 MeV) is used to derive upper and lower bounds for the
  preferred $\Lambda$ value for the dipole case.
\label{fg:chi^2_vs_lambda_dipole}}
\end{center}
\end{figure}

\begin{figure}[*htbp]
\begin{center}
\includegraphics[angle=0, width=0.85\textwidth]{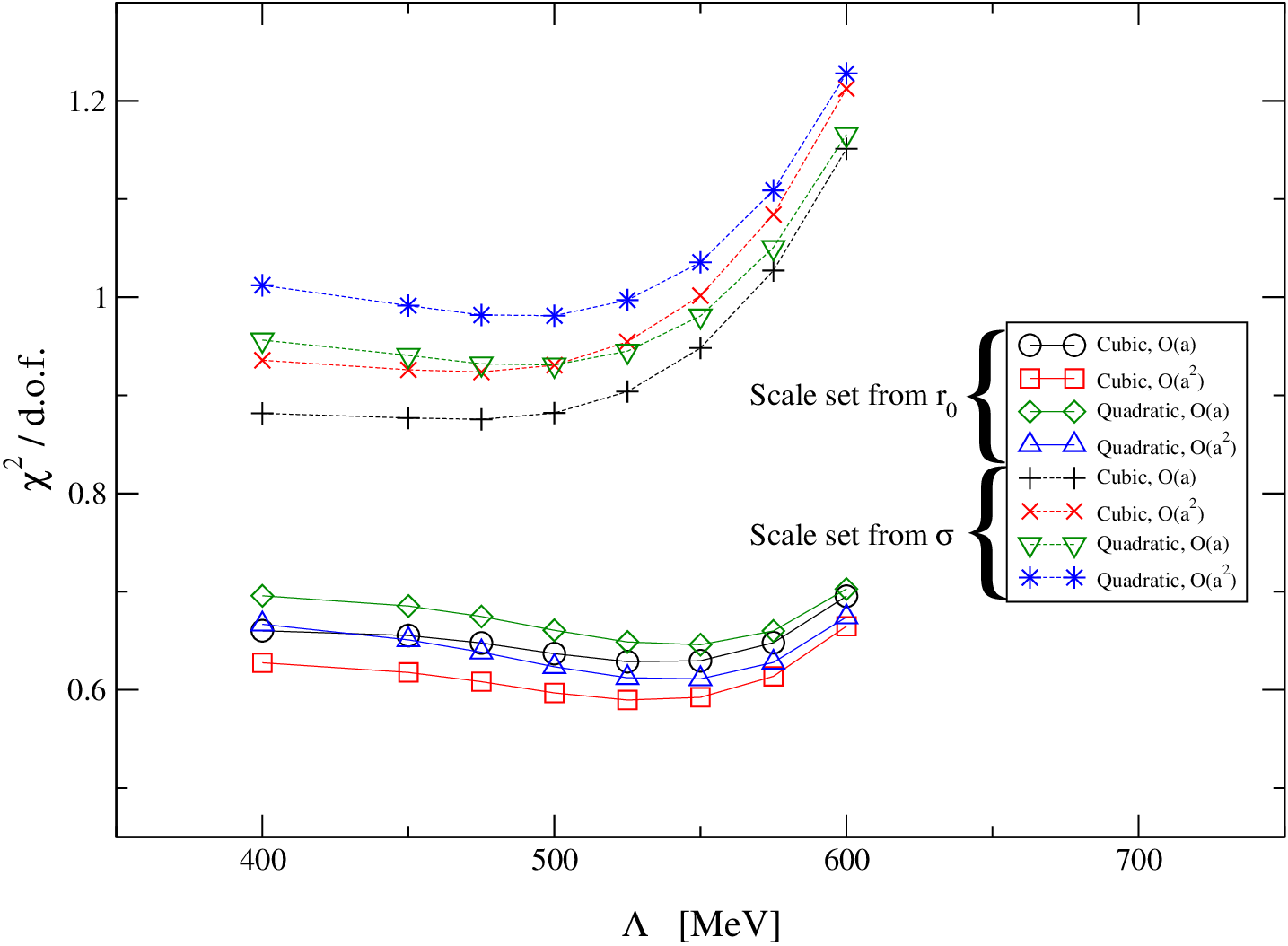}
\caption[A plot of $\chi^2 / d.o.f$ against $\Lambda$ using a gaussian form
factor for the nucleon case.]
{A plot of $\chi^2 / d.o.f$ against $\Lambda$ for the Gaussian form factor.
\label{fg:chi^2_vs_lambda_gaussian}}
\end{center}
\end{figure}

When performing the Adelaide fits, we have to set the $\Lambda$ value,
see Eqs.~(\ref{eq:dipole},\ref{eq:gaussian}).  For numerical reasons,
we chose a set of trial $\Lambda$ values, rather than let it be a free
parameter in the fitting procedure.  In
{}figs.~\ref{fg:chi^2_vs_lambda_dipole} and
\ref{fg:chi^2_vs_lambda_gaussian}, we show the $\chi^2/d.o.f.$ versus
$\Lambda$ for each of the fitting choices in the Adelaide case, for the
dipole and Gaussian form factors, respectively.  This allows us to
study the quality of the fits as a function of $\Lambda$ and to fix
the best value of $\Lambda$ for these fits.


%



%
%


From figs.~\ref{fg:chi^2_vs_lambda_dipole} and
\ref{fg:chi^2_vs_lambda_gaussian}, we see that the behaviour of
$\chi^2$ versus $\Lambda$ is not very dependent on either how the
lattice spacing effects in the $a_0$ coefficient are modelled or on
the order of chiral expansion. Note also that the best $\Lambda$
value, (i.e., the one which minimises $\chi^2$) does not appear to
depend on how the lattice spacing effects are modelled.  Changing
the form factor from dipole to Gaussian results in very similar
$\chi^2$ behaviour, except that the $\Lambda$ value is simply shifted
and the dipole gives a slightly lower $\chi^2$. The biggest effect on the
$\chi^2/d.o.f.$ versus $\Lambda$ curves is the choice of 
whether one uses $r_0$ or
$\sigma$ to set the scale, with $r_0$ clearly producing the best fits.


\begin{table}
\begin{ruledtabular}
\begin{center}
\begin{tabular}{llll}
Form Factor & Chiral     & Lattice & $\Lambda_{\rm{best}}$ \\
            & Expansion  & Spacing & [MeV] \\
\hline
Dipole      & Cubic      & $r_0$      & $594_{-59}^{+32}$ \\
$\;\;\;\;$''&$\;\;\;\;$''& $\sigma$   & 523 \\
&&&\\
$\;\;\;\;$''& Quadratic  & $r_0$      & 600 \\
$\;\;\;\;$''&$\;\;\;\;$''& $\sigma$   & 550 \\
&&&\\
Gaussian    & Cubic      & $r_0$      & 525 \\
$\;\;\;\;$''&$\;\;\;\;$''& $\sigma$   & 475 \\
&&&\\
$\;\;\;\;$''& Quadratic  & $r_0$      & 550 \\
$\;\;\;\;$''&$\;\;\;\;$''& $\sigma$   & 500 \\
\end{tabular}
\end{center}
\end{ruledtabular}
\caption{The best $\Lambda$ values for each of the Adelaide fits from
  figs.~\ref{fg:chi^2_vs_lambda_dipole} and
  \ref{fg:chi^2_vs_lambda_gaussian}. Note that best $\Lambda$ values
  do not depend on whether ${\cal O}(a)$ or ${\cal O}(a^2)$
  corrections in the $a_0$ coefficient are used.
\label{tb:best_lambda}}
\end{table}


For each of the Adelaide fit choices we have determined the ``best''
$\Lambda$ value, i.e., the one which minimises the $\chi^2$. These are
listed in Table~\ref{tb:best_lambda}.  We use these values of
$\Lambda$ to perform the sixteen Adelaide fits in 
Table~\ref{tb:fit_types_nucleon}. The results of these fits, together with
the eight naive fits, are listed in Tables~\ref{tb:global_nucleon_r0} and
\ref{tb:global_nucleon_string}, where the scale is set by $r_{0}$ and
$\sigma$, respectively.


\begin{table}[*htbp]
\begin{ruledtabular}
{\fontsize{10}{10}\selectfont
\begin{center}
\begin{tabular}{ccccccccc}
&&&&&&&&\\
Fit       & Form   & $a_0^{cont}$ &      $X_1$    &      $X_2$    &    $a_2$    &   $a_4$     &    $a_6$    & $\chi^2/d.o.f.$ \\
Approach  & Factor & [GeV]        & [GeVfm$^{-1}$]& [GeVfm$^{-2}$]& [GeV$^{-1}$]& [GeV$^{-3}$]& [GeV$^{-5}$]&                 \\
&&&&&&&&\\
\hline
&&&&&&&&\\
\multicolumn{9}{c}{Cubic chiral extrapolation $\;\;\;\;\;\;$ $a_0$ contains ${\cal O}(a)$} \\
&&&&&&&&\\
Adelaide & dipole  & 1.08\err{ 2}{ 2} & -0.23\err{ 2}{ 3} & - & 1.20\err{ 9}{ 9} &
-0.5\err{ 2}{ 2}   & 0.17\err{ 9}{ 9} & 41 / 75 \\   
Adelaide & Gaussian& 1.08\err{ 2}{ 2} & -0.22\err{ 2}{ 3} & - & 1.19\err{10}{ 9} &
-0.5\err{ 2}{ 2}   & 0.16\err{ 9}{ 9} & 47 / 75 \\
Naive   & -        & 1.02\err{ 2}{ 2} & -0.27\err{ 2}{ 3} & - & 1.29\err{10}{ 9}
& -0.6\err{ 2}{ 2} & 0.21\err{ 9}{ 9} & 45 / 75 \\
&&&&&&&&\\
\hline
&&&&&&&&\\
\multicolumn{9}{c}{Cubic chiral extrapolation $\;\;\;\;\;\;$ $a_0$ contains ${\cal O}(a^2)$} \\
&&&&&&&&\\
Adelaide & dipole  & 1.060\err{14}{16} & - & -0.62\err{ 6}{ 8} & 1.21\err{10}{
8} & -0.53\err{15}{18} & 0.17\err{ 9}{ 8} & 39 / 75 \\
Adelaide & Gaussian& 1.059\err{14}{16} & - & -0.60\err{ 6}{ 8} & 1.19\err{10}{
8} & -0.51\err{15}{18} & 0.16\err{ 9}{ 8} & 44 / 75 \\
Naive   & -        & 0.999\err{13}{17} & - & -0.74\err{ 6}{ 8} & 1.30\err{10}{
8} & -0.64\err{15}{18} & 0.22\err{ 9}{ 8} & 44 / 75\\
&&&&&&&&\\
\hline
&&&&&&&&\\
\multicolumn{9}{c}{Quadratic chiral extrapolation $\;\;\;\;\;\;$ $a_0$ contains ${\cal O}(a)$} \\
&&&&&&&&\\
Adelaide& dipole   & 1.106\err{ 7}{ 8} & -0.23\err{ 2}{ 3} & - & 1.03\err{ 2}{
2} & -0.210\err{15}{17} & - & 43 / 76 \\
Adelaide & Gaussian& 1.101\err{ 7}{ 8} & -0.22\err{ 2}{ 3} & - & 1.03\err{ 2}{
2} & -0.207\err{15}{17} & - & 49 / 76 \\
Naive   & -        & 1.056\err{ 7}{ 8} & -0.28\err{ 2}{ 3} & - & 1.07\err{ 2}{
2} & -0.230\err{15}{17} & - & 49 / 76 \\
&&&&&&&&\\
\hline
&&&&&&&\\
\multicolumn{9}{c}{Quadratic chiral extrapolation $\;\;\;\;\;\;$ $a_0$ contains ${\cal O}(a^2)$} \\
&&&&&&&&\\
Adelaide& dipole   & 1.088\err{ 6}{ 7} & - & -0.63\err{ 6}{ 8} & 1.03\err{ 2}{
2} & -0.209\err{15}{17} & - & 42 / 76 \\
Adelaide& Gaussian & 1.084\err{ 6}{ 7} & - & -0.61\err{ 6}{ 8} & 1.03\err{ 2}{
2} & -0.206\err{15}{17} & - & 47 / 76 \\
Naive   & -        & 1.034\err{ 6}{ 7} & - & -0.75\err{ 6}{ 8} & 1.07\err{ 2}{
2} & -0.230\err{15}{17} & - & 48 / 76 \\
&&&&&&&&\\
\end{tabular}
\end{center}}
\end{ruledtabular}
\caption{The results of the global fit analysis where the scale is set from
$r_{0}$. 
\label{tb:global_nucleon_r0}}

\end{table}



\begin{table}[*htbp]
\begin{ruledtabular}
{\fontsize{10}{10}\selectfont
\begin{center}
\begin{tabular}{ccccccccc}
&&&&&&&&\\
Fit       & Form   & $a_0^{cont}$ &      $X_1$    &      $X_2$    &    $a_2$    &   $a_4$     &    $a_6$    & $\chi^2/d.o.f.$ \\
Approach  & Factor & [GeV]        & [GeVfm$^{-1}$]& [GeVfm$^{-2}$]& [GeV$^{-1}$]& [GeV$^{-3}$]& [GeV$^{-5}$]&                 \\
&&&&&&&&\\
\hline
&&&&&&&&\\
\multicolumn{9}{c}{Cubic chiral extrapolation $\;\;\;\;\;\;$ $a_0$ contains ${\cal O}(a)$} \\
&&&&&&&&\\
Adelaide & dipole  & 1.001\err{15}{14} & -0.18\err{ 2}{ 2} & - & 1.32\err{ 9}{
9} & -0.7\err{ 2}{ 2} & 0.28\err{10}{10} & 61 / 75 \\
Adelaide & Gaussian& 1.002\err{14}{14} & -0.17\err{ 2}{ 2} & - & 1.30\err{ 9}{
8} & -0.7\err{ 2}{ 2} & 0.27\err{10}{10} & 66 / 75 \\
Naive   & -        & 0.966\err{15}{14} & -0.21\err{ 2}{ 2} & - & 1.39\err{ 9}{
9} & -0.8\err{ 2}{ 2} & 0.33\err{10}{10} & 62 / 75 \\
&&&&&&&&\\
\hline
&&&&&&&&\\
\multicolumn{9}{c}{Cubic chiral extrapolation $\;\;\;\;\;\;$ $a_0$ contains ${\cal O}(a^2)$} \\
&&&&&&&&\\
Adelaide & dipole  & 0.986\err{12}{14} & - & -0.48\err{ 6}{ 6} & 1.32\err{ 9}{
8} & -0.7\err{ 2}{ 2} & 0.29\err{10}{ 9} & 66 / 75 \\
Adelaide & Gaussian& 0.988\err{12}{14} & - & -0.44\err{ 6}{ 6} & 1.30\err{ 9}{
8} & -0.7\err{ 2}{ 2} & 0.28\err{10}{ 9} & 69 / 75 \\
Naive   & -        & 0.947\err{13}{14} & - & -0.56\err{ 6}{ 6} & 1.39\err{ 9}{
8} & -0.8\err{ 2}{ 2} & 0.33\err{10}{ 9} & 68 / 75 \\
&&&&&&&&\\
\hline
&&&&&&&&\\
\multicolumn{9}{c}{Quadratic chiral extrapolation $\;\;\;\;\;\;$ $a_0$ contains ${\cal O}(a)$} \\
&&&&&&&&\\
Adelaide & dipole  & 1.036\err{ 7}{ 7} & -0.19\err{ 2}{ 2} & - & 1.08\err{ 2}{
2} & -0.24\err{ 2}{ 2} & - & 67 / 76 \\
Adelaide & Gaussian& 1.036\err{ 7}{ 7} & -0.17\err{ 2}{ 2} & - & 1.08\err{ 2}{
2} & -0.23\err{ 2}{ 2} & - & 71 / 76 \\  
Naive   & -        & 1.006\err{ 6}{ 7} & -0.22\err{ 2}{ 2} & - & 1.11\err{ 2}{
2} & -0.26\err{ 2}{ 2} & - & 69 / 76 \\  
&&&&&&&&\\
\hline
&&&&&&&&\\
\multicolumn{9}{c}{Quadratic chiral extrapolation $\;\;\;\;\;\;$ $a_0$ contains ${\cal O}(a^2)$} \\
&&&&&&&&\\
Adelaide & dipole  & 1.020\err{ 6}{ 6} & - & -0.49\err{ 6}{ 6} & 1.08\err{ 2}{
2} & -0.23\err{ 2}{ 2} & - & 71 / 76 \\
Adelaide & Gaussian& 1.021\err{ 6}{ 6} & - & -0.46\err{ 6}{ 6} & 1.07\err{ 2}{
2} & -0.23\err{ 2}{ 2} & - & 75 / 76 \\
Naive   & -        & 0.987\err{ 6}{ 6} & - & -0.58\err{ 6}{ 6} & 1.11\err{ 2}{
2} & -0.25\err{ 2}{ 2} & - & 75 / 76 \\
&&&&&&&&\\
\end{tabular}
\end{center}
}
\end{ruledtabular}
\caption{The results of the global fit analysis where the scale is set from
$\sigma$. 
\label{tb:global_nucleon_string}}
\end{table}

We summarise the results of Tables~\ref{tb:global_nucleon_r0} and
\ref{tb:global_nucleon_string} below.

\begin{itemize}

\item{\em Fit approach} \\ We see that the smallest $\chi^2 / d.o.f.$
  (indicating the best fitting procedure) is given by the Adelaide
  method which uses a dipole form factor. This has the best $\chi^2 /
  d.o.f. $ in \emph{every} case (independent of how the chiral
  extrapolation was truncated, how the lattice artefacts in $a_0$ were
  modelled and how the spacing was set). Using the Gaussian form
  factor leads to poorer $\chi^2$ values which are similar to, or
  worse than the naive approach.

\item {\em Chiral extrapolation} \\ Errors in the cubic chiral
  coefficient are large compared to their quadratic
  counterparts. However, the cubic fits always produce a non-zero
  $a_{6}$ coefficient and they also lead to a smaller $\chi^2$ value
  than the corresponding quadratic fits. This indicates the need for a
  cubic chiral term. (Note that quadratic chiral fits were preferred
  in the mesonic case~\cite{rhopapers}.)

\item {\em Treatment of the lattice spacing systematics and the fit
  coefficients} \\ We see that the fits with ${\cal O}(a^2)$ rather
  than ${\cal O}(a)$ lead to a lower $\chi^2$ in the $r_0$ case,
  whereas the reverse is true when the string tension is used to set
  the scale. This is another indication that there are ${\cal O}(a)$
  systematics present in the string tension data~\cite{rhopapers}. 

\item{\em Setting the scale} \\ The $\chi^2 / d.o.f.$ using $r_0$ are
  significantly lower than the $\sigma$ fits, independent of 
  whether the Adelaide or naive fits were used. In our study of the
  vector meson~\cite{rhopapers} we found this to be true for the
  Adelaide fits, whereas the naive fits had no preference either way
  for $r_0$ or $\sigma$.

\end{itemize}

From Figs.~\ref{fg:chi^2_vs_lambda_dipole} and
\ref{fg:chi^2_vs_lambda_gaussian} and Tables
\ref{tb:global_nucleon_r0} and \ref{tb:global_nucleon_string}, we see
that the best fit choice is the dipole form factor using the cubic
chiral expansion with ${\cal O}(a^2)$ effects in the $a_0$ coefficient
and using $r_0$ to set the scale. This choice will be used in the next
section to determine the central value of our nucleon mass prediction
and the spread from the other fitting types will be used to define the
error.




\section{Physical predictions}
\label{sec:expt}
In this section we extract a value for the continuum
nucleon mass, $M_N$, in the limit of physical, degenerate quark
masses. 
We also list the renormalised coefficients (i.e. low energy
constants in the chiral expansion of the nucleon mass).
All our predictions will obviously come from the
global fit approach of sec.~\ref{sec:nucleon_global}, since this is a
much more highly constrained method than the alternative method of
sec.~\ref{sec:nucleon_individual}.  We obtain our predictions of $M_N$
by setting $M_{PS}^{deg} = M_{PS}^{non-deg} = M_{PS}^{unit} =
\mu_{\pi}$ in Eqs.~(\ref{eq:nucleon_adel_fit_global},
\ref{eq:nucleon_fit_global} \& \ref{eq:self_terms}) with $\mu_{\pi}$
being the physical pion mass, which we take to be $138$ MeV. We also
set $M_{N}^{deg} = M_{N}^{non-deg}$ and $M_{\Delta}^{deg} =
M_{\Delta}^{non-deg}$ in Eq.~(\ref{eq:self_terms}).  In doing this
we see that the $\eta^{\prime}$ contributions to the total self-energy
(see Eqs.~(\ref{eq:self_heavy_baryon} \& \ref{eq:self_terms})) disappear in
this continuum, physical case as required.
The only remaining term involving $M_{N}$
and $M_{\Delta}$ is the $\sigma^{\pi}_{N \Delta}$ self-energy term
in which we set the mass splitting equal to the physical mass 
splitting of the nucleon
and $\Delta$ (i.e., $293$ MeV~\cite{pdb}).

We make a physical prediction for each different fitting method
(Table~\ref{tb:fit_types_nucleon}) using the coefficients
$(a_0^{cont},~a_2,~a_4~\&~a_6)$ in Tables~\ref{tb:global_nucleon_r0}
and \ref{tb:global_nucleon_string} for the sixteen Adelaide and eight naive
fit cases. For the Adelaide fits, we use the relevant preferred value of
$\Lambda$ taken from Table~\ref{tb:best_lambda}.  We list these
predictions for $M_N$ in Table~\ref{tb:mass_estimates_nucleon}.


\begin{table}[*htbp]
\begin{ruledtabular}
\begin{center}
\begin{tabular}{cccc}
&&&\\
Estimate  & Form   & $M_{N}$ [GeV]        & $M_{N}$ [GeV]         \\
Approach  & Factor & (Scale from $r_{0}$) & (Scale from $\sigma$) \\
&&&\\
\hline
&&&\\
Experimental & -   & \multicolumn{2}{c}{0.939} \\
&&&\\
\hline
\multicolumn{4}{c}{Cubic chiral extrapolation $\;\;\;\;\;\;$ $a_0$ contains ${\cal O}(a)$} \\
&&&\\
Adelaide & dipole   & 0.984\err{15}{15} & 0.950\err{13}{13} \\ 
Adelaide & Gaussian & 0.973\err{15}{15} & 0.938\err{12}{13} \\
Naive    & -        & 1.046\err{15}{15} & 0.992\err{13}{13} \\  
\hline
\multicolumn{4}{c}{Cubic chiral extrapolation $\;\;\;\;\;\;$ $a_0$ contains ${\cal O}(a^2)$} \\
&&&\\
Adelaide & dipole   & 0.965\err{12}{15} & 0.934\err{11}{12} \\
Adelaide & Gaussian & 0.956\err{12}{15} & 0.923\err{11}{12} \\
Naive    & -        & 1.023\err{12}{15} & 0.974\err{11}{12} \\
\hline
\multicolumn{4}{c}{Quadratic chiral extrapolation $\;\;\;\;\;\;$ $a_0$ contains ${\cal O}(a)$} \\
&&&\\
Adelaide & dipole   & 1.006\err{ 7}{ 8} & 0.974\err{ 6}{ 6} \\
Adelaide & Gaussian & 0.986\err{ 7}{ 8} & 0.959\err{ 6}{ 6} \\
Naive    & -        & 1.076\err{ 7}{ 8} & 1.027\err{ 6}{ 6} \\
\hline
\multicolumn{4}{c}{Quadratic chiral extrapolation $\;\;\;\;\;\;$ $a_0$ contains ${\cal O}(a^2)$} \\
&&&\\
Adelaide & dipole   & 0.988\err{ 6}{ 7} & 0.958\err{ 5}{ 6} \\
Adelaide & Gaussian & 0.969\err{ 6}{ 7} & 0.945\err{ 5}{ 6} \\
Naive    & -        & 1.054\err{ 6}{ 7} & 1.008\err{ 5}{ 6} \\
\end{tabular}
\end{center}
\end{ruledtabular}
\caption[Estimates of $M_{N}$ obtained from the global fits.]
        {Estimates of $M_{N}$ obtained from the global fits. (The
          errors are statistical only.) Our experimental estimate
          comes from a simple average of the proton and neutron
          masses.
\label{tb:mass_estimates_nucleon}}
\end{table}

\begin{figure}[*htbp]
\begin{center}
\includegraphics[angle=0, width=0.85\textwidth]{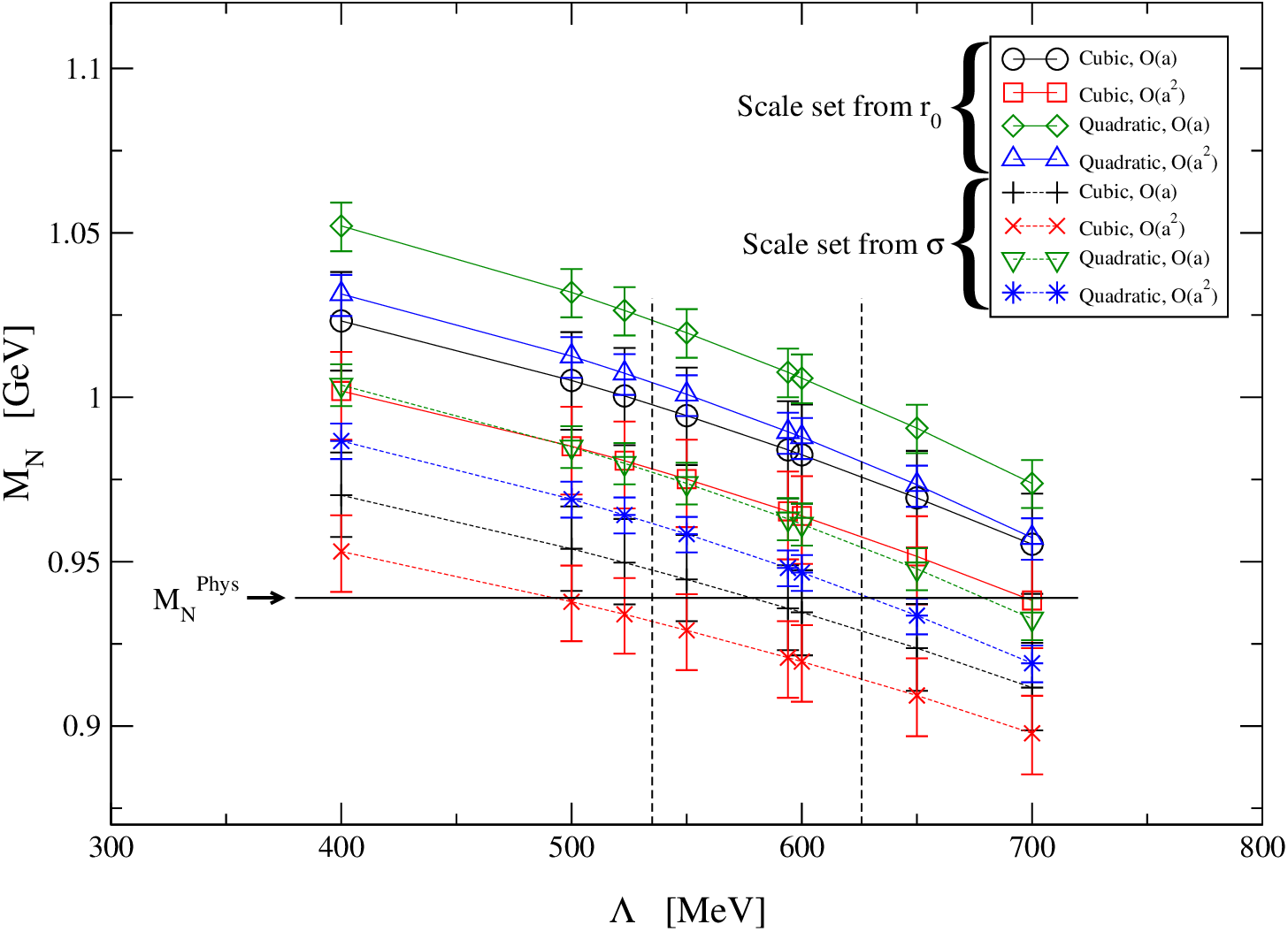}
\caption[A plot of $M_{N}$ as a function of $\Lambda$ from the Adelaide approach using a
dipole form factor.]
{A plot of $M_{N}$ as a function of $\Lambda$ from the Adelaide approach using a
dipole form factor.
Recall that the best $\Lambda$ value when the scale is set from $r_0$
for the dipole form factor is $\Lambda = 594$ MeV.
The two vertical dashed lines define the range of acceptable $\Lambda$
values (535 MeV $\le \Lambda \le$ 626 MeV) obtained by increasing
$\chi^2$ by unity in Fig.~\ref{fg:chi^2_vs_lambda_dipole}.
\label{fg:nucleon_mass_vs_lambda_dipole}}
\end{center}
\end{figure}
\begin{figure}[*htbp]
\begin{center}
\includegraphics[angle=0,
width=0.85\textwidth]{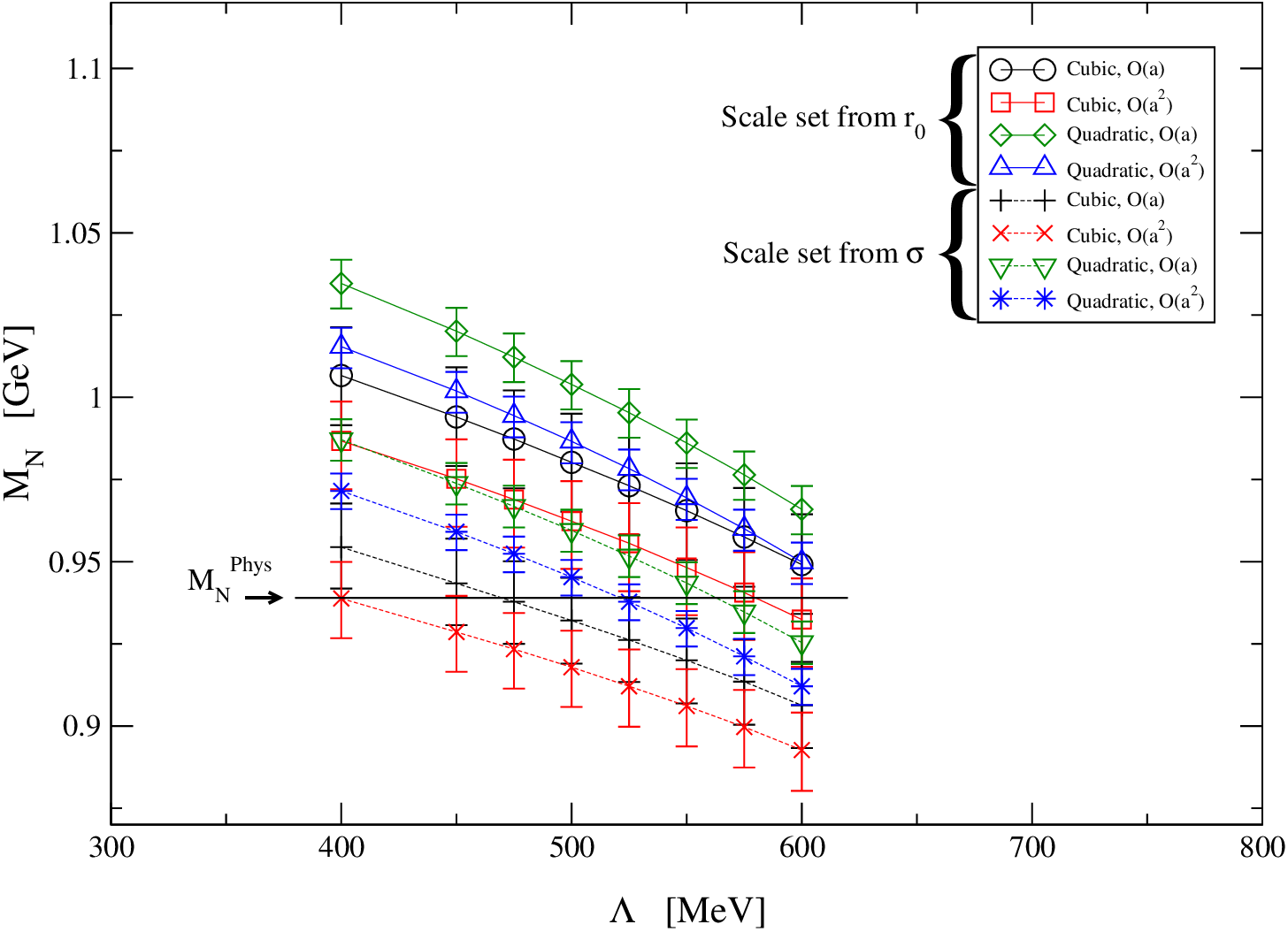}
\caption[A plot of $M_{N}$ as a function of $\Lambda$ from the Adelaide approach using a
Gaussian form factor.]
{A plot of $M_{N}$ as a function of $\Lambda$ from the Adelaide approach using a
Gaussian form factor.
Recall that the best $\Lambda$ value when the scale is set from $r_0$
for the Gaussian form factor is $\Lambda = 525$ MeV.
\label{fg:nucleon_mass_vs_lambda_gaussian}}
\end{center}
\end{figure}

In Figs.~\ref{fg:nucleon_mass_vs_lambda_dipole} and
\ref{fg:nucleon_mass_vs_lambda_gaussian} we present a graphical
representation of our study of the $\Lambda$ dependence of $M_{N}$ for
both the dipole and Gaussian form factors.  For the dipole case, we
include the acceptable range for the $\Lambda$ parameter which is
represented by two vertical dashed lines. This range is defined from
our plots of $\chi^{2} / d.o.f$ against $\Lambda$
({}Figs.~\ref{fg:chi^2_vs_lambda_dipole} \& \ref{fg:chi^2_vs_lambda_gaussian})
by increasing $\chi^{2}$ from its minimum by unity.


We summarise the results of this section which are outlined in Table
\ref{tb:mass_estimates_nucleon} and figures
\ref{fg:nucleon_mass_vs_lambda_dipole} and
\ref{fg:nucleon_mass_vs_lambda_gaussian} below. 

\begin{itemize}

\item The statistical errors in the mass estimates are typically less
  than 1\% for the quadratic extrapolations and less than 2\% for the
  cubic extrapolations.

\item We see disagreement between all types of fit when different
  methods are used to set the scale. When the scale is set from
  $r_{0}$ the mass predictions are always higher than when the scale
  is set from $\sigma$.

\item We also see that the $M_N$ predictions within one particular method
  (i.e. the Adelaide dipole, Adelaide Gaussian or naive method) have a
  variation in the results of between 3\% and 5\%, with the largest
  variation in the naive mass predictions. This disagreement suggests
  some instability in the fits, possibly because the lattice
  systematics are more complicated than we have assumed.

\item The Adelaide method always produces a mass prediction closer
  to the physical nucleon mass. For the cubic fits the Adelaide mass
  predictions are very accurate compared to their naive counterparts.
  They are typically within two statistical error bars of the
  experimental mass.

\item The variation of $M_{N}$ in the region of allowed values of
  $\Lambda$ is very small for each different fit. Typically of the
  order of the other uncertainties.

\end{itemize}

As with the results of the meson study~\cite{rhopapers} we
conclude by noting all of these points favour the Adelaide approach
over the naive method.  This suggests that the Adelaide method should be
the preferred method when performing chiral extrapolations and is a
significant improvement over the naive method.  To give the final
value for $M_{N}$ for both the Adelaide method and the naive method,
we use our preferred choices: the cubic chiral extrapolation with
${\cal O}(a^{2})$ corrections in $a_{0}$; $r_{0}$ to set the scale;
and, for the Adelaide method we use the dipole form factor. We quote
an error that is based on the spread in the mass predictions (for the
$r_{0}$ case only). We also (for the Adelaide method) include an
estimate of the error associated with the $\Lambda$ parameter which is
taken from the vertical dashed lines in 
{}Fig.~\ref{fg:nucleon_mass_vs_lambda_dipole}.

Hence our final mass estimate for the nucleon is
\bea\label{eq:mass_nucleon_final_adel}
M_{N}^{Adelaide} &=& 965(15)\er{41}{0}\er{13}{8} \textrm{MeV} \\
\label{eq:mass_nucleon_final_naive}
M_{N}^{Naive}    &=& 1023(15)\er{53}{0} \textrm{MeV}
\eea
where the first error is statistical and the second is taken from the
fit procedure. The third error in the Adelaide case is due to the
$\Lambda$ parameter. We have not considered any error that may be
associated with the determination of $r_{0}$ which we take to be
0.49 fm.  We note that the
Adelaide central value would be $1\sigma$ (of the combined in
quadrature errors) away from experiment by simply rescaling $r_0$
upwards from 0.49 fm by around 1\%.  This corresponds exactly with
what we found in the $\rho-$meson mass case~\cite{rhopapers}.
However, other results have hinted at smaller values of $r_0$
\cite{Aubin:2004wf, Khan:2006de, Boucaud:2007uk}.
To a good approximation, the $M_N$ estimates 
for these values of $r_0$ near 0.49fm can be obtained by a simple scaling
of the values in Eqs.(\ref{eq:mass_nucleon_final_adel} \&
\ref{eq:mass_nucleon_final_naive}).

We now turn to the renormalised coefficients, $c_{0,2}$, which are
defined from Eqs.(\ref{eq:self_heavy_baryon} \&
\ref{eq:nucleon_adel_fit_form}).
\bea
 \nonumber
M_N &=& a_0                    + a_2(M_{PS}^{deg})^2                    + a_4(M_{PS}^{deg})^4 + \sigma_{NN}^{\pi} + \sigma_{N\Delta}^{\pi} + \ldots \\ \nonumber
    &\equiv& a_0               + a_2(M_{PS}^{deg})^2                    + a_4(M_{PS}^{deg})^4 \\ \nonumber
    &&+ (a_{NN}^{\pi})^{(0)}     + (a_{NN}^{\pi})^{(2)}(M_{PS}^{deg})^2     + c_{LNA}(M_{PS}^{deg})^3 \\ \nonumber
    &&+ (a_{N\Delta}^{\pi})^{(0)} + (a_{N\Delta}^{\pi})^{(2)}(M_{PS}^{deg})^2 + c_{NLNA}(M_{PS}^{deg})^4 \ln(M_{PS}^{deg}) + \ldots \\ \nonumber
    &\equiv& c_0               + c_2(M_{PS}^{deg})^2                     + c_4(M_{PS}^{deg})^4 \\
    &&                         + c_{LNA}(M_{PS}^{deg})^3                  + c_{NLNA}(M_{PS}^{deg})^4\ln(M_{PS}^{deg}) + \ldots
\label{eq:c_02}
\eea

Again we have used the fact that the $\eta^{\prime}$ terms,
$\sigma_{NN}^{\eta^{\prime}}$ and $\sigma_{N\Delta}^{\eta^{\prime}}$,
disappear in the continuum, physical case, and we note that the
$\sigma_{NN}^{\pi}$,$\sigma_{N\Delta}^{\pi}$ contributions reproduce
the LNA and NLNA terms respectively once they are chirally
expanded. By expanding $\sigma_{NN}^{\pi}$ and
$\sigma_{N\Delta}^{\pi}$ (from eq.(\ref{eq:self_terms})) about
$M_{PS}^2=0$ we obtain the values for $(a_{NN}^{\pi})^{(0,2)}$ and
$(a_{N\Delta}^{\pi})^{(0,2)}$ listed in Tables \ref{tb:c_02_r0} and
\ref{tb:c_02_string}. Note that these values correspond to the fit
procedures used in Tables \ref{tb:global_nucleon_r0} and
\ref{tb:global_nucleon_string}.  The $\Lambda$ values used are those
in Table \ref{tb:best_lambda}. From eq(\ref{eq:c_02}) we have
\be
c_{0,2} = a_{0,2} + (a_{NN}^{\pi})^{(0,2)} + (a_{N\Delta}^{\pi})^{(0,2)}.
\label{eq:c_02defn}
\ee
Tables \ref{tb:c_02_r0} and \ref{tb:c_02_string} list the values
of $c_{0,2}$ for each of the fit procedures.
Using the same preferred fitting method as in the nucleon mass case
(cubic chiral extrapolation with ${\cal O}(a^2)$ corrections in $a_0$
with $r_0$ to set the scale and using the dipole form factor) we
obtain
\be \label{eq:c02}
c_0 = 0.930(16)\mbox{\err{42}{12}} \mbox{GeV} \;\;\;\;\;\;\;\;\;\;\;\;\;\;\;\;\;\;
c_2 = 2.61(10)\mbox{\err{17}{17}}  \mbox{GeV}^{-1},
\ee
where the first error is statistical and the second is from the fit
procedure. 
The $c_0$ value is very close to the experimental value of $M_N$ as
expected. 
Furthermore, the $c_2$ value compares very favourably with
\cite{Leinweber:2003dg} who obtained $c_2 = 2.80(33)(35)$ GeV$^{-1}$
(where we
note the statistical error in \cite{Leinweber:2003dg} was reported at
$2\sigma$).
Note however that this was with full, 2-flavour QCD (rather
than pQQCD); the reduced error in our case is due to the larger dataset
generically available in pQQCD. Other work, \cite{Meissner:2006kf},
calculates a similar quantity they call $c_1$ from
low energy $\pi-N$ fits. This is related to our $c_2$ via
\be \label{eq:c1}
c_1 \equiv -c_2/4.
\ee
\cite{Meissner:2006kf} obtains $c_1 = -0.9$\err{0.5}{0.2} GeV$^{-1}$,
which, using
eq.(\ref{eq:c1}), predicts $c_2 = 3.6$\err{0.8}{2.0} GeV$^{-1}$.
Again, this estimate
is consistent with our result in eq(\ref{eq:c02}), but with
significantly larger errors. Note however, that the result from
\cite{Meissner:2006kf} uses experimental (2+1 flavour) data.  Thus it
is clear that there is significant benefit in simulating pQQCD since
it generates a larger dataset which results in a corresponding
reduction in the errors associated with physical predictions and low
energy constants.


\begin{table}[*htbp]
\begin{ruledtabular}
{\fontsize{10}{10}\selectfont
\begin{center}
\begin{tabular}{ccccccc}
&&&&&&\\
Form   &
$(a_{NN}^{\pi})^{(0)}$  & $(a_{N\Delta}^{\pi})^{(0)}$ &$c_0$ &
$(a_{NN}^{\pi})^{(2)}$  & $(a_{N\Delta}^{\pi})^{(2)}$ & $c_2$ \\
Factor &
[GeV]                 & [GeV]                      & [GeV] &
[GeV$^{-1}$]           & [GeV$^{-1}$]               & [GeV$^{-1}$] \\
&&&&&&\\
\hline
&&&&&&\\
\multicolumn{7}{c}{Cubic chiral extrapolation $\;\;\;\;\;\;$ $a_0$ contains ${\cal O}(a)$} \\
&&&&&&\\
dipole  &
-0.074  & -0.056 & 0.95\err{ 2}{ 2} &
 1.056  &  0.339 & 2.60\err{ 9}{ 9} \\   
Gaussian&
-0.082  & -0.059 & 0.94\err{ 2}{ 2} &
 1.191  &  0.401 & 2.78\err{10}{ 9} \\
&&&&&&\\
\hline
&&&&&&\\
\multicolumn{7}{c}{Cubic chiral extrapolation $\;\;\;\;\;\;$ $a_0$ contains ${\cal O}(a^2)$} \\
&&&&&&\\
dipole  &
-0.074  & -0.056 & 0.930\err{14}{16} &
 1.056  &  0.339 & 2.61\err{10}{8} \\
Gaussian&
-0.082  & -0.059 & 0.918\err{14}{16} &
 1.191  &  0.401 & 2.78\err{10}{ 8} \\
&&&&&&\\
\hline
&&&&&&\\
\multicolumn{7}{c}{Quadratic chiral extrapolation $\;\;\;\;\;\;$ $a_0$ contains ${\cal O}(a)$} \\
&&&&&&\\
dipole  &
-0.077  & -0.058 & 0.972\err{ 7}{ 8} &
 1.066  &  0.345 & 2.44\err{ 2}{2} \\
Gaussian&
-0.094  & -0.069 & 0.938\err{ 7}{ 8} &
 1.248  &  0.433 & 2.71\err{ 2}{2} \\
&&&&&&\\
\hline
&&&&&&\\
\multicolumn{7}{c}{Quadratic chiral extrapolation $\;\;\;\;\;\;$ $a_0$ contains ${\cal O}(a^2)$} \\
&&&&&&\\
dipole  &
-0.077  & -0.058 & 0.954\err{ 6}{ 7} &
 1.066  &  0.345 & 2.44\err{ 2}{2} \\
Gaussian &
-0.094  & -0.069 & 0.921\err{ 6}{ 7} &
 1.248  &  0.433 & 2.71\err{ 2}{2} \\
&&&&&&\\
\end{tabular}
\end{center}}
\end{ruledtabular}
\caption{The renormalised coefficients $c_{0,2}$ using the scale set
  from $r_0$.
\label{tb:c_02_r0}}

\end{table}


\begin{table}[*htbp]
\begin{ruledtabular}
{\fontsize{10}{10}\selectfont
\begin{center}
\begin{tabular}{ccccccc}
&&&&&&\\
Form   &
$(a_{NN}^{\pi})^{(0)}$  & $(a_{N\Delta}^{\pi})^{(0)}$ &$c_0$ &
$(a_{NN}^{\pi})^{(2)}$  & $(a_{N\Delta}^{\pi})^{(2)}$ & $c_2$ \\
Factor &
[GeV]                 & [GeV]                      & [GeV] &
[GeV$^{-1}$]           & [GeV$^{-1}$]               & [GeV$^{-1}$] \\
&&&&&&\\
\hline
&&&&&&\\
\multicolumn{7}{c}{Cubic chiral extrapolation $\;\;\;\;\;\;$ $a_0$ contains ${\cal O}(a)$} \\
&&&&&&\\
dipole  &
-0.051  & -0.036  & 0.914\err{15}{14}   &
 0.929  &  0.273  & 2.52\err{ 9}{9}    \\
Gaussian&
-0.061  & -0.042  & 0.900\err{14}{14}   &
 1.078  &  0.338  & 2.72\err{ 9}{8}     \\
&&&&&&\\
\hline
&&&&&&\\
\multicolumn{7}{c}{Cubic chiral extrapolation $\;\;\;\;\;\;$ $a_0$ contains ${\cal O}(a^2)$} \\
&&&&&&\\
dipole  &
-0.051  & -0.036  & 0.899\err{12}{14}   &
 0.929  &  0.273  & 2.52\err{ 9}{8}     \\
Gaussian&
-0.061  & -0.042  & 0.886\err{12}{14}   &
 1.078  &  0.338  & 2.72\err{ 9}{8}     \\
&&&&&&\\
\hline
&&&&&&\\
\multicolumn{7}{c}{Quadratic chiral extrapolation $\;\;\;\;\;\;$ $a_0$ contains ${\cal O}(a)$} \\
&&&&&&\\
dipole  &
-0.059  & -0.043  & 0.934\err{ 7}{ 7}   &
 0.977  &  0.298  & 2.36\err{ 2}{2}     \\
Gaussian&
-0.071  & -0.050  & 0.915\err{ 7}{ 7}   &
 1.134  &  0.369  & 2.58\err{ 2}{2}      \\
&&&&&&\\
\hline
&&&&&&\\
\multicolumn{7}{c}{Quadratic chiral extrapolation $\;\;\;\;\;\;$ $a_0$ contains ${\cal O}(a^2)$} \\
&&&&&&\\
dipole  &
-0.059  & -0.043  & 0.918\err{ 6}{ 6}   &
 0.977  &  0.298  & 2.36\err{ 2}{2}      \\
Gaussian&
-0.071  & -0.050  & 0.900\err{ 6}{ 6}   &
 1.134  &  0.369  & 2.57\err{ 2}{2}     \\
&&&&&&\\
\end{tabular}
\end{center}
}
\end{ruledtabular}
\caption{The renormalised coefficients $c_{0,2}$ using the scale set
  from $\sigma$. 
\label{tb:c_02_string}}
\end{table}




\section{Conclusions}
\label{sec:nucleon_cons}
While the computing resources needed to generate gauge configurations 
with dynamical fermions are very significant, the computation of 
hadron properties with different valence quark masses is relatively low cost
for any given sea quark mass. Thus, if one can deal with the effects of 
partial quenching in a controlled manner, this approach offers a potentially 
cost effective way to increase the statistical precision of the 
final, physical results. This philosophy has already been successfully 
applied to the mass of the $\rho$ meson in earlier work~\cite{rhopapers}.
Here, we have estimated the nucleon mass and its corresponding low energy
constants from a large CP-PACS
simulation~\cite{cppacs} of partially-quenched baryon masses using the
Adelaide chiral extrapolation approach. We have shown that this method
is a valid approach to the study of the chiral properties of the
nucleon mass in the partially-quenched theory.  As a comparison, we
have also used a naive (polynomial) fitting procedure, but found that
it is a much poorer predictor of the experimental number compared with
the Adelaide approach.  Our predictions from both methods are given in
Eqs.~(\ref{eq:mass_nucleon_final_adel} \&
\ref{eq:mass_nucleon_final_naive}).  

We have shown that a single global fit of all 80 degenerate nucleon
mass points from~\cite{cppacs} at different valence and sea quark
masses, and at different lattice spacings is possible using the
Adelaide approach. This fit procedure has only 5 fit parameters and
includes chiral effects, finite volume and finite lattice spacing
effects. The only systematic deviation from nature not included is the
number of sea quark flavours which is two in the CP-PACS simulation.
The Adelaide method includes a form factor and an associated scale,
$\Lambda$. In this work, we have shown that both the type of form
factor, and the preferred value of $\Lambda$ can be determined.  As a
by-product, we have been able to determine the low energy constant in
the chiral expansion of the nucleon mass, $c_2$, to a remarkable level
of accuracy (see eq.(\ref{eq:c02})). Again this emphasises the benefits
that can be obtained from the larger dataset that pQQCD data affords.



\section*{Acknowledgements}
CRA and WA would like to thank the CSSM for their support and kind
hospitality.  WA would like to thank PPARC for travel support.  The
authors would like to thank Stewart Wright for helpful comments.
This work is supported by the Australian Research Council and by U.S.
DOE Contract No. DE-AC05-06OR23177, under which Jefferson Science
Associates, LLC operates Jefferson Laboratory, and DE-AC02-06CH11357,
under which UChicago Argonne, LLC operates Argonne National
Laboratory.




\end{document}